\documentclass[a4paper,11pt]{article}

\usepackage{jheppub} 

\usepackage[T1]{fontenc} 
\usepackage{amsmath,amssymb,braket}
\usepackage{graphicx}
\usepackage[utf8]{inputenc}
\usepackage{color}

\preprint{%
  \begin{flushright}
    UTHEP-708 \\ UTCCS-P-107 \\ KANAZAWA-17-11
  \end{flushright}
}

\title{Tensor network formulation for two-dimensional lattice $\mathcal{N}=1$ Wess--Zumino model}

\author[a]{Daisuke Kadoh,}
\author[b,c]{Yoshinobu Kuramashi,}
\author[c]{Yoshifumi Nakamura,}
\author[d]{Ryo Sakai,}
\author[d]{\\ Shinji Takeda,}
\author[b]{and Yusuke Yoshimura}

\affiliation[a]{Research and Educational Center for Natural Sciences, Keio University, Yokohama 223-8521, Japan}
\affiliation[b]{Center for Computational Sciences, University of Tsukuba, Tsukuba 305-8577, Japan}
\affiliation[c]{RIKEN Advanced Institute for Computational Science, Kobe 650-0047, Japan}
\affiliation[d]{Institute for Theoretical Physics, Kanazawa University, Kanazawa 920-1192, Japan}

\emailAdd{kadoh@keio.jp}
\emailAdd{kuramasi@het.ph.tsukuba.ac.jp}
\emailAdd{nakamura@riken.jp}
\emailAdd{sakai@hep.s.kanazawa-u.ac.jp}
\emailAdd{takeda@hep.s.kanazawa-u.ac.jp}
\emailAdd{yoshimur@ccs.tsukuba.ac.jp}

\abstract{%
  Supersymmetric models with spontaneous supersymmetry breaking suffer from the notorious sign problem in stochastic approaches.
  By contrast, the tensor network approaches do not have such a problem since they are based on deterministic procedures.
  In this work, we present a tensor network formulation of the two-dimensional lattice $\mathcal{N}=1$ Wess--Zumino model while showing that numerical results agree with the exact solutions for the free case.
}

\begin{document} 
\maketitle
\flushbottom

%
%
%
%

\section{Introduction}
\label{sec:Introduction}

Supersymmetric field theories have attracted great attention because they provide a deep insight 
about the non-perturbative physics~\cite{Seiberg:1994rs, Seiberg:1994aj, Seiberg:1994pq} and have a close relation with 
the gravitational theory~\cite{Maldacena:1997re}.
The lattice simulations are promising approaches to obtain a further understanding of them. 
However, it is generally difficult to use the standard Monte Carlo techniques for the lattice supersymmetric theories 
on account of the sign problem, and the theories with the supersymmetry breaking may be the most difficult 
cases as suggested from the vanishing Witten index~\cite{Witten:1982df}. 
In this paper, we apply the tensor network approach, which is free of the sign problem, to the two-dimensional 
lattice $\mathcal{N}=1$ Wess--Zumino model in order to make a breakthrough on the issue.

The two-dimensional $\mathcal{N}=1$ Wess--Zumino model is a supersymmetric theory 
in which a real scalar interacts with a Majorana fermion via the Yukawa term
originate from the superpotential~\cite{Wess:1974tw}.
The supersymmetry is spontaneously broken for the supersymmetric $\phi^4$ theory in a finite volume~\cite{Witten:1982df},
and the Witten index becomes zero because the fermion Pfaffian has both the positive and negative signs.
For the infinite volume case, the absence of the non-renormalization theorem suggests that the breaking may occur 
even at the perturbative level~\cite{Bartels:1983wm, Synatschke:2009nm},
and the theory has a rich phase structure which should be clarified 
by numerical methods free from the sign problem.

Although the lattice regularization generally breaks the $\mathcal{N}=1$ supersymmetry for the interacting theories
in contrast to the case of $\mathcal{N}=2$ model~\cite{
  Cecotti:1982ad,
  Sakai:1983dg,
  Kikukawa:2002as,
  Giedt:2004qs,
  Kadoh:2010ca},\footnote{
  Non-local formulations of the Wess--Zumino model have been studied in refs.~\cite{
    Dondi:1976tx,
    Kadoh:2009sp,
    Asaka:2016cxm,
    DAdda:2017bzo}
}
it is known that the breaking term caused by the lattice cut-off disappears in the continuum limit 
for an appropriate lattice action at least in the perturbation theory~\cite{Golterman:1988ta}.
In the action, the Wilson terms are included in both the fermion and the boson sectors,
so that the supersymmetry is exactly realized in the free-theory limit.
Some numerical studies have been already done 
in the low-dimensional Wess--Zumino model~\cite{
  Beccaria:1998vi,
  Catterall:2001fr,
  Beccaria:2004pa,
  Giedt:2005ae,
  Bergner:2007pu,
  Kastner:2008zc,
  Kawai:2010yj,
  Wozar:2011gu,
  Steinhauer:2014yaa}.   
In our study we use the tensor network approach to investigate the $\mathcal{N}=1$ supersymmetric model much deeper.

The tensor renormalization group (TRG) is a coarse-graining algorithm for tensor networks, which is based on the singular value decomposition (SVD).
The TRG was originally introduced in a two-dimensional classical spin model~\cite{Levin:2006jai}.
Since the TRG was extended to the Grassmann TRG for models including Grassmann variables~\cite{Gu:2010yh,Gu:2013gba},
some studies of fermionic systems have been reported so far.
In two-dimensional quantum field theories, it was already applied to the lattice $\phi^{4}$ theory~\cite{Shimizu:2012wfa}
and to the lattice Schwinger model~\cite{Shimizu:2014uva,Shimizu:2014fsa} 
and the lattice $N_{\mathrm{f}}=1$ Gross--Neveu model~\cite{Takeda:2014vwa}, which are Dirac fermion systems.
For the lattice $\mathcal{N}=1$ Wess--Zumino model, we have to clarify a method to construct a tensor network representation
for the Majorana fermions with the Yukawa-type interaction and for the case of next-nearest-neighbor interacting bosons
which originate from the Wilson term.

In this paper, we show that the partition function of the lattice $\mathcal{N}=1$ Wess--Zumino model can be expressed as a tensor network
for any superpotential and any value of the Wilson parameter $r$.
Refining the known method for the Dirac fermions~\cite{Takeda:2014vwa}, we present a way of making a tensor network representation
for Majorana fermions.
For the boson action, we can change it to one with up to nearest-neighbor interactions by introducing two auxiliary fields.
Then we also show a tensor network representation for bosons with a new discretization scheme. 
In order to test our formulation, we compute the Witten index by using the Grassmann TRG.
Although we give a method of constructing tensors for any interacting case,
in numerical test we devote ourselves to the free Wess--Zumino model,
which is the most suitable test bed for a tensor network representation.
This is because non-trivial structures of tensor arise from the hopping terms in the lattice action.
This point will be discussed along with the details of the tensor network representation in the main part of this paper.
The computation is done with $r=1/\sqrt{2}$, so that one of the two auxiliary fields is decoupled
to reduce the computational cost.

This paper is organized as follows.
We first recall the two-dimensional $\mathcal{N}=1$ Wess--Zumino model and its lattice version with the detailed notations in section~\ref{sec:Model}.
In section~\ref{sec:TensorNetwork}, tensor network representation for the fermion part and the boson part are individually constructed. 
By combining those two results, the tensor network representation for the total partition function is also given.
Section~\ref{sec:Results} shows the numerical results for the free case, and we compare them with the exact ones.
A summary and a future outlook are given in section~\ref{sec:Summary}.

%
%
%
%

\section{Two-dimensional $\mathcal{N}=1$ Wess--Zumino model}
\label{sec:Model}

%
%

\subsection{Continuum theory}

Two-dimensional $\mathcal{N}=1$ Wess--Zumino model is a supersymmetric theory
that consists of a real scalar field $\phi\left(x\right)$ and a Majorana fermion field $\psi\left(x\right)$.
In the Euclidean space-time, the corresponding action is given by 
\begin{align} 
  \label{eq:2}
  S_{\mathrm{cont.}} 
  = \int \mathrm{d}^{2}x \left\{
  \frac{1}{2}\left(\partial_{\mu}\phi\right)^{2}
  + \frac{1}{2} W^\prime \left(\phi\right)^{2} 
  + \frac{1}{2}\bar{\psi} \left(\gamma_\mu \partial_\mu
  + W^{\prime\prime}\left(\phi\right)\right)\psi
  \right\},
\end{align}
where $\gamma_\mu$ is the gamma matrix which satisfies
\begin{align}
  \label{eq:25}
  \left\{ \gamma_{\mu}, \gamma_{\nu} \right\} =2\delta_{\mu \nu}, \qquad \gamma_\mu=\gamma_\mu^\dag.
\end{align}
The Lorentz index $\mu$ takes two values $1$ or $2$, and the Einstein summation convention 
is used throughout this paper.  Showing the indices in the spinor space explicitly, 
$\gamma_\mu$ and $\psi\left(x\right)$ are written as $\left(\gamma_\mu\right)_{\alpha\beta}$ and 
$\psi_\alpha\left(x\right)$ for $\alpha, \beta=1,2$.
The spinor index $\alpha$ and the space-time coordinate $x$ are often suppressed without notice.
$W\left(\phi\right)$ is an arbitrary real function of $\phi$, which is referred to as the superpotential in the superfield formalism,  
and gives the Yukawa- and $\phi^n$-type interactions 
with common coupling constants. 
$W^{\prime}\left(\phi\right)$  is the first differential of 
$W\left(\phi\right)$ with respect to $\phi$, that is, $W^{\prime}\left(\phi\right) \equiv \left(\mathrm{d}/\mathrm{d}\phi\right) W\left(\phi\right)$.

The Majorana fermion $\psi$ satisfies
\begin{align}
  \label{eq:majorana}
  \bar{\psi} =-\psi^{\mathrm{T}} C^{-1}, 
\end{align}
where $C$ is  the charge conjugation matrix which obeys
\begin{align}
  \label{eq:24}
  C^{\mathrm{T}}=-C, \quad C^{\dagger}=C^{-1}, \quad C^{-1}\gamma_{\mu}C=-\gamma_{\mu}^{\mathrm{T}}.
\end{align}
For any $W\left(\phi\right)$, the action in eq.~\eqref{eq:2} is invariant under the supersymmetry transformation
\begin{align}
  \label{eq:susy}
  &\delta \phi\left(x\right)
    = \bar{\epsilon} \psi\left(x\right), \\
  &\delta \psi\left(x\right)
    = \left(\gamma_\mu \partial_\mu \phi\left(x\right) - W^{\prime}\left(\phi\left(x\right)\right)\right) \epsilon,
\end{align}
where $\epsilon$ is a global Grassmann parameter with two components
and $\bar \epsilon$ satisfies eq.~\eqref{eq:majorana}.

%
%

\subsection{Lattice theory}
\label{sec:lattice_theory}

Let us consider a two-dimensional square lattice with the lattice spacing $a$ 
and  the volume $V=aN_1\times aN_2$, where $N_1, N_2 \in \mathbb{N}$. 
In this paper, $a$ is set to unity, and the lattice sites are simply expressed by integers:
\begin{align}
  \Gamma = \Set{(n_1,n_2)| n_\mu=1,2,\ldots,N_\mu \ \ \text{for } \mu=1,2}.
\end{align}
All of the fields live on the lattice sites $n \in \Gamma$ 
and satisfy the periodic boundary conditions in both directions. 
The forward and the backward difference operators, $\partial_{\mu}$ and $\partial^{*}_{\mu}$, are given by
\begin{align}
  \label{eq:6}
  &\partial_{\mu} \phi_{n}
    = \phi_{n+\hat{\mu}} - \phi_{n},\\
  &\partial_{\mu}^{*}\phi_{n}
    = \phi_{n} - \phi_{n-\hat{\mu}},
\end{align}
where $\hat \mu$ is the unit vector along the $\mu$-direction,
and the symmetric difference operator 
is given by $ \partial_{\mu}^{\mathrm{S}} = \left( \partial_\mu + \partial_\mu^*  \right)/2$.

We define the lattice Wess--Zumino model according to ref.~\cite{Golterman:1988ta}:
\begin{align}
  \label{eq:3}
  S
  = \sum_{n\in\Gamma}\left\{ \frac{1}{2}\left(\partial_{\mu}^{\mathrm{S}}\phi_n\right)^{2} 
  + \frac{1}{2}\left( W^\prime \left(\phi_{n}\right) -\frac{r}{2} \partial_{\mu}^{\mathrm{}} \partial_{\mu}^{*} \phi_{n} \right)^{2} 
  + \frac{1}{2}\bar{\psi}_{n} D\psi_{n} \right\},
\end{align}
where 
the lattice Dirac operator $D$ which acts as $D\psi_m=D_{mn}\psi_n$ is given by
\begin{align}
  \label{eq:4}
  D_{mn}
  = \left( \gamma_\mu \partial_\mu^{\mathrm{S}}
  - \frac{r}{2}\partial_{\mu}\partial_{\mu}^{*}\right)_{mn}
  + W^{\prime\prime} \left(\phi_n\right)\delta_{mn}
\end{align}
with the nonzero real Wilson parameter $r$.
In the following, $S_{\rm B}$ denotes the pure boson part of the action:
\begin{align}
  \label{eq:aa}
  S_{\rm B}
  = \sum_{n\in\Gamma}\left\{ \frac{1}{2}\left(\partial_{\mu}^{\mathrm{S}}\phi_n\right)^{2} 
  + \frac{1}{2}\left( W^\prime \left(\phi_{n}\right) -\frac{r}{2} \partial_{\mu}^{\mathrm{}} \partial_{\mu}^{*} \phi_{n} \right)^{2} 
  \right\}.
\end{align}
Note that the kinetic term of $\phi$ is given by the symmetric difference operator
instead of the forward one in the naive boson action
\begin{align}
  \label{eq:SBnaive}
  S_{\rm B, naive} 
  = \sum_{n\in\Gamma}\left\{ \frac{1}{2}\left(\partial_{\mu}\phi_n\right)^{2} 
  + \frac{1}{2}\left( W^\prime \left(\phi_{n}\right)  \right)^{2} 
  \right\},
\end{align}
and an extra Wilson term is included in the boson sector. 
In this paper we refer to the bosons with the Wilson term as the Wilson bosons 
in the same sense as the Wilson fermions.
While $S_{\mathrm{B, naive}}$ has only the nearest-neighbor interactions, 
$S_{\rm B}$ has the next-nearest-neighbor ones that cause difficulties 
in constructing the tensor network representation of the partition function.
This point will be discussed later.

In the free theory with
\begin{align}
  W\left(\phi\right) = \frac{1}{2}m\phi^2,
  \label{eq:free_P}
\end{align}
the action in eq.~\eqref{eq:3} is invariant under a lattice version of the supersymmetry transformation
\begin{align}
  \label{eq:5}
  &\delta \phi_{n}
    = \bar{\epsilon}\psi_{n}, \\
  &\delta \psi_{n}
    = \left\{ \gamma_{\mu} \partial^{\mathrm{S}}_{\mu}\phi_{n} + \frac{r}{2}\left(\partial_{\mu} \partial_{\mu}^* \phi\right)_{n} - W^\prime \left(\phi_{n}\right)\right\} \epsilon
\end{align}
even at a finite lattice spacing because $S_{\rm B}$ has the similar structure 
with the Wilson--Dirac operator $D$ in eq.~\eqref{eq:4} in contrast to the naive one.
For the interacting cases, however, the invariance is explicitly broken owing to the lack of the Leibniz rule 
for the lattice difference operators. 
The broken supersymmetry is shown to be restored in the continuum limit, at least,
at all orders of the perturbation~\cite{Golterman:1988ta}.

The associated partition function is defined in the usual manner:
\begin{align}
  \label{eq:Z}
  Z=\int \mathcal{D}\phi \mathcal{D}\psi  e^{-S}
\end{align}
with the path integral measures
\begin{align}
  \label{eq:phi_int}
  & \int \mathcal{D}\phi \equiv \prod_{n\in \Gamma} \int_{-\infty}^{\infty}  \frac{\mathrm{d}\phi_n}{\sqrt{2\pi}}, 
  \\
  & \int \mathcal{D}\psi \equiv \prod_{n\in \Gamma} \int \mathrm{d}\psi_{n,1} \mathrm{d}\psi_{n,2}.
\end{align}
Here $\mathrm{d}\psi_{n,\alpha}$ is a measure of the Grassmann integral defined in the following.
The Grassmann variable
$\xi_i$ and its measure $\mathrm{d}\xi_i$ ($i=1,\ldots, I$) satisfy
\begin{align}
  \label{eq:grassman_number}
  \left\{
  \xi_i,\xi_j
  \right\}=\left\{
  \xi_i,\mathrm{d}\xi_j
  \right\}=\left\{
  \mathrm{d}\xi_i,\mathrm{d}\xi_j
  \right\}=0
  && \text{for all } i,j.
\end{align}
The Grassmann integral is then defined by
\begin{align}
  \label{eq:grassman_integral}
  \int \mathrm{d}\xi_i 1=0,
  \qquad
  \int \mathrm{d}\xi_i \xi_i=1
  &&\text{for }i=1,2,\ldots,I,
\end{align}
which suggests that $\int \mathrm{d}\xi_i$ is equivalent to $\partial/\partial \xi_i$.

In the free theory, the boson and the fermion are decoupled from each other,
and the respective partition functions are given by
\begin{align}
  \label{eq:Z_B_exact}
  & Z_{\rm B, exact}
    =\prod_{p_1,p_2} \frac{1}{ \sqrt{ \sum_{\mu=1}^2 \sin^2 p_\mu 
    + \left(m+2r\sum_{\mu=1}^2 \sin^2\left(p_\mu/2\right) \right)^2}},
  \\
  \label{eq:Z_F_exact}
  & Z_{\rm F, exact} 
    =\frac{{\rm sign}\left\{m\left(m+4r\right)\right\}}{Z_{\mathrm{B, exact}}},
\end{align}
where $p_\mu=2\pi n/N_\mu$ ($n=0,1,2,\ldots,N_\mu-1$)
and the product in eq.~\eqref{eq:Z_B_exact} is taken for all possible momenta~\cite{Wolff:2007ip}. 
Note that $Z_{\rm B}=\infty$ ($Z_{\rm F}=0$) for $m=0,-2r,-4r$ when $N_\mu$ is an even integer
because the first term and the second term in the square root in eq.~\eqref{eq:Z_B_exact} simultaneously vanish for certain combinations of $p_{1}$ and $p_{2}$.
Thus we find that the Witten index, which is defined as the partition function with periodic boundary conditions in a finite volume,
\begin{align}
  \label{eq:W_exact}
  Z_{\mathrm{exact}}={\rm sign}\left\{m\left(m+4r\right)\right\}
\end{align}
reproduces the continuum one, ${\rm sign}\left\{m\right\} $, for $\left|m\right| \ll 1$.

After integrating the fermion field, the partition function can also be written as
\begin{align}
  \label{eq:9}
  Z=\int \mathcal{D}\phi e^{-S_{\rm B}} \mathrm{Pf}\left(C^*D\right),
\end{align}
where the Pfaffian of a $2I\times 2I$ anti-symmetric matrix $A$ is defined by
\begin{align}
  \mathrm{Pf} (A)
  = \int \mathrm{d}\xi_1 \mathrm{d}\xi_2\cdots \mathrm{d}\xi_{2I} e^{-\frac{1}{2} \xi_{i} A_{ij}\xi_j}
\end{align}
for Grassmann variables $\left\{\xi_i\right\}$ and corresponding measures $\left\{\mathrm{d}\xi_i\right\}$.
The fermion Pfaffian $\mathrm{Pf}\left(C^*D\right)$ flips its sign depending on the scalar field in the interacting cases.
To overcome this sign problem we employ the TRG method, whose first step is to represent
eq.~\eqref{eq:Z} as a network of uniform tensors, which is explained in the next section.

%
%
%
%

\section{Tensor network representation of partition function}
\label{sec:TensorNetwork}

%
%

\subsection{Fermion Pfaffian}
\label{fermion_sector}

We construct a tensor network representation for the fermion part of eq.~\eqref{eq:Z}
\begin{align}
  \label{eq:ZF}
  Z_{\rm F} = \int\mathcal{D}\psi e^{-\frac{1}{2}\sum_{n\in \Gamma}\bar \psi_n D\psi_{n}},
\end{align}
which yields the Pfaffian after integrating the fermion field as found in eq.~\eqref{eq:9}.
The basic idea follows from refs.~\cite{Shimizu:2014uva,Takeda:2014vwa} which deal with the Dirac fermions.   
We describe the procedure for the Majorana fermions with any value of the Wilson parameter $r$.

Now we use the following representations for $\gamma_\mu$ and $C$ that satisfy eqs.~\eqref{eq:25} and~\eqref{eq:24}:
\begin{align}
  \label{eq:1}
  \gamma_{1}= \sigma_1,
  \qquad 
  \gamma_{2}=\sigma_3,
  \qquad C=-i\sigma_2,         
\end{align}
where $\sigma_i$ is the standard Pauli matrix.
The method presented in this section is applicable to any possible choice of $\gamma_\mu$ and $C$, 
and they just lead to different tensors. 
Then the Majorana spinor takes the form
\begin{align}
  \label{eq:majorana_spinor}
  \psi_{n}= 
  \begin{pmatrix}
    \psi_{n,1}  \\
    \psi_{n,2} 
  \end{pmatrix}
  ,\qquad \bar\psi_n=
  \begin{pmatrix}
    \psi_{n,2}, & -\psi_{n,1} 
  \end{pmatrix}
                  ,
\end{align}
and we obtain 
\begin{align}
  \label{eq:10}
  - \frac{1}{2}\sum_{n\in \Gamma } \bar \psi_{n} D\psi_{n}
  =&\sum_{n\in \Gamma } \bigg\{
     \left(\frac{1+r}{2}\right)
     \left( \tilde \psi_{n+\hat 1,2} \tilde \psi_{n,1} +\psi_{n+\hat 2,2} \psi_{n,1} \right)
     \nonumber\\
   &+ \left(\frac{1-r}{2}\right) 
     \left( \tilde \psi_{n+\hat 1,1} \tilde \psi_{n,2} +\psi_{n+\hat 2,1} \psi_{n,2} \right)
     + \left(W^{\prime\prime}\left(\phi_n\right)+2r\right) \psi_{n,1} \psi_{n,2}
     \bigg\},
\end{align}
where 
\begin{align}
  & \tilde \psi_{n,1}=\frac{1}{\sqrt{2}}\left( \psi_{n,2}+ \psi_{n,1}  \right), \\
  & \tilde \psi_{n,2}=\frac{1}{\sqrt{2}}\left(\psi_{n,2} - \psi_{n,1}  \right),
\end{align}
which are local transformations of the field variable $\psi_n$.
$\tilde \psi_{n,\alpha}$ is introduced only to write eq.~\eqref{eq:10} as simple as possible.
Note that the second term in eq.~\eqref{eq:10} disappears for $r=1$ because the hopping terms 
in eq.~\eqref{eq:4} are proportional to the projection operators, $\left(1\pm\gamma_\mu\right)/2$.

Let us expand the four types of hopping factors in eq.~\eqref{eq:ZF}:
\begin{align}
  \label{eq:11}
  &\hspace{-3.2em}e^{-\frac{1}{2}\sum_{n\in \Gamma}\bar \psi_{n} D \psi_{n}}
    \nonumber\\
  =\prod_{n\in \Gamma} \Bigg\{
  &\sum_{u_n=0}^1  \left( \frac{1+r}{2}  \tilde \psi_{n+\hat 1,2} \tilde \psi_{n,1} \right)^{u_n}
    \sum_{v_n=0}^1  \left( \frac{1-r}{2}  \tilde \psi_{n+\hat 1,1} \tilde \psi_{n,2} \right)^{v_n}
    \nonumber\\
  &\cdot \sum_{p_n=0}^1  \left( \frac{1+r}{2}  \psi_{n+\hat 2,2} \psi_{n,1} \right)^{p_n}
    \sum_{q_n=0}^1  \left( \frac{1-r}{2}  \psi_{n+\hat 2,1} \psi_{n,2} \right)^{q_n}
    e^{(W^{\prime\prime}\left(\phi_n)+2r\right) \psi_{n,1} \psi_{n,2}}
    \Bigg\}.
\end{align}
We will see that $u_n,v_n,p_n,q_n$, which take 0 or 1 because of the nilpotency of $\psi_{n,\alpha}$ (and $\tilde \psi_{n,\alpha}$), 
are regarded as the indices of tensors. 
The four types of hopping factors have the same structure as $\Psi_{n+\hat\mu} \Phi_{n}$,
where $\Psi_{n+\hat\mu}$ and $\Phi_{n}$ are single-component Grassmann numbers. 
It is straightforward to show 
\begin{align}
  \label{eq:key_identity}
  \Psi_{n+\hat\mu} \Phi_{n} = \int \left(\Psi_{n+\hat\mu}\mathrm{d}\bar\theta_{n+\hat \mu}\right)\left(\Phi_{n}\mathrm{d}\theta_n\right)\left(\bar\theta_{n+\hat\mu}\theta_n\right),
\end{align}
where new independent Grassmann numbers $\theta_n$, $\bar\theta_{n+\hat\mu}$ 
and the corresponding measures $\mathrm{d}\theta_n$, $\mathrm{d}\bar\theta_{n+\hat\mu}$ 
satisfy
eqs.~\eqref{eq:grassman_number} and~\eqref{eq:grassman_integral} with the periodic boundary conditions.
By applying this identity to each hopping factor in eq.~\eqref{eq:11} individually,
one can make a tensor network representation.

Then the fermion part of the partition function is represented as a product of tensors
\begin{align}
  \label{eq:TNR_F}
  Z_{\rm F}=\sum_{\left\{u,v,p,q\right\}} &\prod_{n \in \Gamma} 
                                            {T_{\rm F}}\left(\phi_n\right)_{u_n v_n p_n q_n u_{n-\hat 1} v_{n-\hat 1} p_{n-\hat 2} q_{n-\hat 2}} \nonumber \\
                                          & \cdot \int {\cal D} \Xi_{uvpq} 
                                            \prod_{n \in \Gamma} 
                                            \left(\bar \xi_{n+\hat 1}\xi_n\right)^{u_n}  \left(\bar \chi_{n+\hat 1}\chi_n\right)^{v_n}  \left(\bar \eta_{n+\hat 2}\eta_n\right)^{p_n}  \left(\bar \zeta_{n+\hat 2}\zeta_n\right)^{q_n}
\end{align}
with
\begin{align}
  \label{eq:measures}
  {\cal D} \Xi_{uvpq} 
  =\prod_{n \in \Gamma} 
  \mathrm{d}\xi^{u_n}_n \mathrm{d}\chi^{v_n}_n \mathrm{d}\eta^{p_n}_n \mathrm{d}\zeta^{q_n}_n 
  \mathrm{d} \bar \xi^{u_{n-\hat 1}}_n \mathrm{d}\bar \chi^{v_{n-\hat 1}}_n \mathrm{d} \bar \eta^{p_{n-\hat 2}}_{n} \mathrm{d} \bar \zeta^{q_{n-\hat 2}}_{n},
\end{align}
where $\xi_n, {\bar \xi}_n, \chi_n, {\bar \chi}_n, \eta_n, {\bar \eta}_n, \zeta_n, {\bar \zeta}_n$,
and those with bars are single-component Grassmann numbers introduced in the manner of eq.~\eqref{eq:key_identity},
and $\sum_{\{u,v,\cdots\}}$ means the summation of all possible configurations of the indices: $\prod_{n \in \Gamma} \left(\sum_{u_n=0}^1 \sum_{v_n=0}^1 \cdots \right)$.
The new Grassmann numbers and their corresponding measures
satisfy the same anti-commutation relations and boundary conditions as those of the original ones.
The tensor $T_{\rm F}$ is defined as
\begin{align}
  \label{eq:fermion_tensor}
  {T_{\rm F}}\left(\phi\right)_{u v p q a b c d} 
  = 
  \int\mathrm{d} \Psi \mathrm{d} \Phi
  &e^{(W^{\prime\prime}\left(\phi\right)+2r) \Psi \Phi} 
    \left\{
    \Psi^{d}
    \Phi^{c} 
    \tilde\Psi^{b} 
    \tilde\Phi^{a}  
    \Phi^{q}
    \Psi^{p} 
    \tilde\Phi^{v}
    \tilde\Psi^{u} 
    \right\} 
    \nonumber \\
  &\cdot
    \left(\sqrt{\frac{1+r}{2}}\right)^{u+p+a+c} \left(\sqrt{\frac{1-r}{2}}\right)^{v+q+b+d}
\end{align}
for all possible indices
with single-component Grassmann numbers $\Psi$, $\Phi$,
$\tilde\Psi = \left(\Phi+\Psi\right)/\sqrt{2}$, $\tilde\Phi = \left(\Phi-\Psi\right)/\sqrt{2}$.
By integrating $\Psi$ and $\Phi$
by hand, we can obtain the tensor elements.
In the case of $r=1$, note that the indices $v_n$ and $q_n$ vanish and that the Grassmann fields $\chi_n$ and $\zeta_n$ are decoupled
because the second term in the RHS of eq.~\eqref{eq:10} is absent.
In that case, the tensor network representation becomes much simpler:
\begin{align}
  \left. Z_{\rm F} \right|_{r=1}=\sum_{\left\{u,p\right\}} \prod_{n \in \Gamma} 
  {T_{\rm F}\left(\phi_n\right)}_{u_n p_n u_{n-\hat 1} p_{n-\hat 2}}  \int \mathrm{d}\xi^{u_n}_n  \mathrm{d}\eta^{p_n}_n 
  \mathrm{d} \bar \xi^{u_{n-\hat 1}}_n \mathrm{d} \bar \eta^{p_{n-\hat 2}}_{n} 
  \prod_{n \in \Gamma} 
  \left(\bar \xi_{n+\hat 1}\xi_n\right)^{u_n}   \left(\bar \eta_{n+\hat 2}\eta_n\right)^{p_n}, 
\end{align}
where 
\begin{align}
  {T_{\rm F}}\left(\phi\right)_{ijkl} 
  = 
  \int\mathrm{d} \Psi \mathrm{d} \Phi
  e^{\left(W''\left(\phi\right)+2\right) \Psi \Phi} 
  \Phi^l  
  \tilde\Phi^k  
  \Psi^j 
  \tilde\Psi^i .
\end{align}

It is rather straightforward to show that eq.~\eqref{eq:11} is reproduced
from eq.~\eqref{eq:TNR_F} with eqs.~\eqref{eq:measures} and~\eqref{eq:fermion_tensor}
and from the identity in eq.~\eqref{eq:key_identity}.
We now note that the eight Grassmann measures in the RHS of eq.~({\ref{eq:measures}}) 
should be in this order and that the set of measures at the site $n$ commutes with ones at different lattice sites
because they are Grassmann-even as a set 
for non-zero elements of the tensor given in eq.~\eqref{eq:fermion_tensor}.

The indices $x_n \equiv (u_n, v_n)$ and the Grassmann fields $\xi_n, \chi_n$ carry 
the information of the hopping factors with $\mu=1$ 
as indicated by the last factors in eq.~\eqref{eq:TNR_F} while $t_n \equiv (p_n, q_n)$ and  $\eta_n, \zeta_n$ 
are related to the hopping with $\mu=2$.
In this sense,  $x_n$, $t_n$, $x_{n-\hat 1}$,  $t_{n-\hat 2}$, which are the indices of the tensor in eq.~\eqref{eq:TNR_F}, 
can be interpreted as being defined on  the four links which stem from the site $n$.
Since each index is shared by two tensors which are placed on the nearest-neighbor lattice sites (see eq.~\eqref{eq:TNR_F}), 
we can find that the partition function $Z_{\rm F}$ is expressed as a network of the tensor
${T_{\rm F}}_{x_n t_n x_{n-\hat 1} t_{n-\hat 2}}$ on the two-dimensional square lattice $\Gamma$.

If one uses another representation of $\gamma_\mu$ and $C$, then the same partition function is given by a different tensor.
This means that the tensor network representation is not uniquely determined.

%
%

\subsection{Boson partition function}
\label{sec:Boson_Z}

The tensor network representation is also constructed  for the pure boson part of eq.~\eqref{eq:Z}
\begin{align}
  \label{eq:ZB}
  Z_{\rm B}=\int \mathcal{D}\phi e^{-S_{\rm B}}
\end{align}
with $S_{\rm B}$ in eq.~\eqref{eq:aa}.
It is, however, not straightforward to construct a simple representation
because $S_{\rm B}$ has the next-nearest-neighbor interactions
and $\phi$ is a non-compact field. 
A popular way to avoid the former issue is to rewrite $S_{\rm B}$ 
in a nearest-neighbor form with the aid of auxiliary fields.
For the latter, we employ a new method using a discretization for the integrals of $\phi$.\footnote{
  A method for treating the non-compact field using a discretization is already proposed 
  in the pioneering work by Y. Shimizu~\cite{Shimizu:2012wfa}.
  We thank him for pointing out a new idea~\cite{Shimizu:private} presented in this paper. 
}
After these procedures, we find that a discretized version of eq.~\eqref{eq:ZB} can be expressed 
as a tensor network for arbitrary discretization schemes.

Since the formulation is actually irrelevant to the details of the scalar theory,
we will derive a tensor network for a general theory:
\begin{align}
  \label{eq:ZB_varphi}
  Z_{\rm B}=\int \mathcal{D}\varphi e^{- \tilde S_{\rm B}\left(\varphi\right)},
\end{align}
where $\int \mathcal{D} \varphi= \int_{-\infty}^\infty\prod_{n \in \Gamma } \mathrm{d}\varphi_{n, 1} \mathrm{d}\varphi_{n, 2}\cdots \mathrm{d}\varphi_{n, N}$.
We assume that $\tilde S_{\rm B}(\varphi)$ is invariant under the PT-transformation 
on a two-dimensional square lattice and has the interactions up to the nearest-neighbor, 
and that $\varphi_n$ is a non-compact real field with $N$ components.
As seen in section~\ref{sec:remarks}, it is very easy to extend it to the non PT-symmetric case.

We will show that eq.~\eqref{eq:ZB} can be expressed in the form of eq.~\eqref{eq:ZB_varphi} with $N=3$ 
in section~\ref{sec:auxiliary_field}.
After decomposing the hopping terms of $\tilde S_{\rm B}$ in section~\ref{sec:symmetric}
and introducing a formal discretization for the integrals of $\varphi$ in section~\ref{sec:discretization},
we give the tensor network representation for a discretized version of eq.~\eqref{eq:ZB_varphi} in section~\ref{sec:decomposition}.

%
%

\subsubsection{Introduction of auxiliary fields}
\label{sec:auxiliary_field}

The boson action $S_{\rm B}$ in eq.~\eqref{eq:aa} is transformed 
into a nearest-neighbor form using two real auxiliary fields $G$ and $H$:
\begin{align}
  \label{eq:ZB_newS}
  Z_{\rm B}=\int \mathcal{D}\phi  \mathcal{D}G \mathcal{D}H e^{- \tilde S_{\rm B}},
\end{align}
where 
\begin{align}
  \label{eq:7}
  \tilde S_{\mathrm{B}}
  = S_{\mathrm{B, naive}}
  + \frac{1}{2}\sum_{n \in \Gamma}
  \big\{
  &G_{n}^{2} + H_{n}^{2}
    -\left( r W' \left(\phi_{n}\right)+\alpha G_n+\beta H_n\right)
    \left(\phi_{n+\hat 1} +\phi_{n-\hat 1}- 2\phi_n\right) \nonumber\\
  &-\left( r W' \left(\phi_{n}\right)+\alpha G_n-\beta H_n\right)
    \left(\phi_{n+\hat 2} +\phi_{n-\hat 2}- 2\phi_n\right) 
    \big\}
\end{align}
with $S_{\mathrm{B,naive}}$ given in eq.~\eqref{eq:SBnaive}, $\alpha=\sqrt{(1-2r^2)/2}$, and $\beta=1/\sqrt{{2}}$. 
Note that
$\alpha$ is real for $\left|r\right| \le 1/\sqrt{2}$ but becomes a pure imaginary for $\left|r\right|>1/\sqrt{2}$.
The integral measures for $G_n$ and  $H_n$ are defined in exactly the same way as $\phi_n$ in eq.~\eqref{eq:phi_int}.
Although, in general, two auxiliary fields are necessary for the next-nearest-neighbor interactions in two directions, 
it is somewhat surprising to find that $G$ is decoupled from the other fields for particular values $r=\pm 1/\sqrt{2}$,
and the required auxiliary field turns out to be only $H$.

It is clear that $\tilde S_{\rm B}$ has only the on-site and the nearest-neighbor interactions 
which are invariant under the PT-transformation
\begin{align}
  \phi_n,\ H_n,\ G_n \quad \rightarrow \quad \phi_{-n},\ H_{-n},\ G_{-n}.
\end{align}
Defining a three-component field variable
\begin{align}
  \label{eq:simple_variables}
  \varphi_n = \left(\varphi_{n1},\varphi_{n2},\varphi_{n3}\right) = \left(\frac{\phi_n}{\sqrt{2\pi}}, \frac{H_n}{\sqrt{2\pi}},\frac{G_n}{\sqrt{2\pi}}\right),
\end{align}
we find that eq.~\eqref{eq:ZB_newS} is just eq.~\eqref{eq:ZB_varphi} with $N=3$.

%
%

\subsubsection{Symmetric property of local Boltzmann weight}
\label{sec:symmetric}

In the previous section~\ref{sec:auxiliary_field}, we found that eq.~\eqref{eq:ZB} is a special case of eq.~\eqref{eq:ZB_varphi}. 
Hereafter we will try to derive a tensor network representation of general one~\eqref{eq:ZB_varphi}.
Before that, let us see the hopping structure of the local Boltzmann weight, 
which is an important building block of the tensor as shown in following sections~\ref{sec:discretization} and~\ref{sec:decomposition}.

It can be easily shown that $\tilde S_{\rm B}$ is expressed as 
\begin{align}
  \label{eq:S_twoL}
  \tilde S_{\rm B}=\sum_{n \in \Gamma} L_1\left(\varphi_n,\varphi_{n+\hat 1}\right) + \sum_{n \in \Gamma} L_2\left(\varphi_n,\varphi_{n+\hat 2}\right),
\end{align}
where $L_\mu$ is symmetric in the sense that $L_\mu\left(\varphi,\varphi^{\prime}\right)=L_\mu\left(\varphi^{\prime},\varphi\right)$
which is a consequence of the PT-invariance of the action.\footnote{
  We can express the action as $\tilde{S}_{\mathrm{B}} = \sum_{n \in \Gamma} \sum_{\mu = 1}^{2} K_{\mu}\left(\varphi_{n}, \varphi_{n+\hat{\mu}}\right)$ using a trial choice of $K_{\mu}$.
  Actually, $K_{\mu}\left(\varphi_{n}, \varphi_{n+\hat{\mu}}\right)$ transforms to $K_{\mu}\left(\varphi_{-n}, \varphi_{-n-\hat{\mu}}\right)$ by the PT-transformation,
  and the PT-invariance of the action tells us that $\tilde{S}_{\mathrm{B}} = \sum_{n \in \Gamma} \sum_{\mu = 1}^{2} K_{\mu}\left(\varphi_{n+\hat{\mu}}, \varphi_{n}\right)$.
  Thus the symmetric $L_{\mu}$ is always defined as $L_{\mu}\left(\varphi, \varphi^{\prime}\right) = \left(K_{\mu}\left(\varphi, \varphi^{\prime}\right) + K_{\mu}\left(\varphi^{\prime}, \varphi\right)\right)/2$.
}
All of the hopping terms with respect to the $\mu$-direction are in $L_\mu\left(\varphi_n,\varphi_{n+\hat\mu}\right)$.
This decomposition is actually not unique because the positions of the on-site interactions 
and some constants are free to choose.

For our case in eq.~\eqref{eq:7}, we find
\begin{align}
  \label{eq:Lmuourcase}
  L_\mu\left(\varphi_n,\varphi_{m}\right)  = &\frac{1}{2} \left(\phi_m-\phi_n\right)^2   
                                               + \frac{1}{8} \left(
                                               W^\prime\left(\phi_n\right)^2
                                               + G_{n}^{2} + H_{n}^{2}
                                               + W^{\prime}\left(\phi_{m}\right)^{2}
                                               + G_{m}^{2} + H_{m}^{2}
                                               \right)
                                               \nonumber\\
                                             &\begin{aligned}[t]
                                               -\frac{1}{2} \Bigl( &r W^\prime \left(\phi_{n}\right)  + \alpha G_n + \left(-1\right)^{\delta_{\mu2}}\beta H_{n} \\
                                               &- r W^\prime \left(\phi_{m}\right)  - \alpha G_m - \left(-1\right)^{\delta_{\mu2}}\beta H_{m} \Bigr) \left(\phi_m -\phi_n\right).
                                             \end{aligned}\nonumber\\[-4.8ex]
\end{align}
Note that $\beta H_n$ and $\beta H_{m}$ have the different signs for $\mu=2$.

The Boltzmann factor $e^{- \tilde S_{\rm B}}$ can be written as 
\begin{align}
  \label{eq:16}
  e^{-\tilde S_{\mathrm{B}}}
  = \prod_{n \in \Gamma} \prod_{\mu = 1}^{2} 
  f_{\mu}\left(\varphi_{n}, \varphi_{n+\hat{\mu}}\right)
\end{align}
with
\begin{align}
  \label{eq:f}
  f_{\mu}\left(\varphi,\varphi' \right)=e^{-L_\mu\left(\varphi,\varphi' \right)},
\end{align}
which is symmetric in the same sense as that of $L_\mu$.
This symmetric property plays an important role in the subsequent discussion.

%
%

\subsubsection{Discretization of non-compact field}
\label{sec:discretization}

The non-compactness of the variable $\varphi$ is cumbersome 
in extracting the tensor structure from $f_\mu\left(\varphi_{n},\varphi_{n+\hat{\mu}} \right)$ in practice. 
There are several possible ways to make the indices of the tensor. 
In our method, we first carry out a discretization of the variable $\varphi$ itself, 
which automatically makes the partition function in eq.~\eqref{eq:ZB_varphi} into a discretized form.

To make the discussion of the discretization clearly understood, 
let us begin with a one-dimensional integral
\begin{align}
  \label{eq:one_dim_int}
  I= \int_{-\infty}^{\infty} {\mathrm{d}x} f\left(x\right),
\end{align}
which converges for a given function $f\left(x\right)$. 
We can formally approximate this integral with a discretized form
\begin{align}
  \label{eq:one_dim_sum}
  I\left(K\right)= {\sum_{x \in S_{K}}}^{\!\!\left(\mathrm{disc.}\right)} f\left(x\right),
\end{align}
for which 
$I=\lim_{K \rightarrow \infty} I\left(K\right)$
is simply assumed.
$K$ is a parameter to control the approximation of the integral by the sum.
Now we suppose that  $S_K$ is a set containing $K$ numbers, $x_1, x_2, \ldots, x_K$, 
which are given by a discretization scheme $\left(\mathrm{disc.}\right)$, 
and that $\sum_{x \in S_{K}}^{\left(\mathrm{disc.}\right)}$ is a summation of $x \in S_K$ with some factors, 
for instance,  $\left(\mathrm{disc.}\right)$-dependent weights.  
A multi-dimensional extension ($S_K\rightarrow S_K^N$) is straightforward 
by defining $S_K^N$ as a set of the multi-dimensional discrete points.

The Gauss--Hermite quadrature gives a concrete example of this abstract definition. 
The RHS of eq.~\eqref{eq:one_dim_sum} is then defined as follows:
\begin{align}
  \label{eq:GH-quardrature}
  {\sum_{x \in S_{K}}}^{\!\!\left(\mathrm{GH}\right)}  f\left(x\right)
  \equiv \sum_{i=1}^{K} w_i e^{x_{i}^{2}} f\left(x_{i}\right).
\end{align}
Here
$x_i$ $(i=1,\ldots,K)$ is the $i$-th root of the $K$-th Hermite polynomial, and 
$w_i$ are the weights given by the Hermite polynomial and $x_i$.
The RHS of  eq.~\eqref{eq:GH-quardrature} has an extra exponential function 
because this quadrature is designed so that $f(x)$ which has a damping factor $e^{-x^2}$ is well approximated.
In this case, we find that $S_K$ is a set of the roots and the weight $w_i e^{x_i^2}$ is the ingredient of $\sum^{\left(\mathrm{GH}\right)}$.      
For a well-behaved $f\left(x\right)$, one can expect that $I=\lim_{K \rightarrow \infty} I\left(K\right)$.

With the prescriptions above,  eq.~\eqref{eq:ZB_varphi} can be discretized as
\begin{align}
  \label{eq:26}
  Z_{\mathrm{B}}\left(K\right)
  = \prod_{n \in \Gamma} \left(\frac{1}{\sqrt{2\pi}}\right)^{N} {\sum_{\varphi_n \in S_{K}^N}}^{\!\!\!\!\left(\mathrm{disc.}\right)}
  \prod_{\mu=1}^{2} f_{\mu}\left(\varphi_{n}, \varphi_{n+\hat{\mu}}\right)
\end{align}
by replacing the measures for $\varphi_n$ by $\sum_{\varphi_n \in S_{K}^N}^{\left(\mathrm{disc.}\right)}$.\footnote{
  Here a one-dimensional discretization is applied to each component of $\varphi_{n}$.
  We may also use a more general scheme that cannot be
  written as the superposition of one-dimensional discretization.
}~\footnote{
  In general one can set different discrete points for each direction: $K_{i} \neq K_{j}$ for $i \neq j$,
  although in the following we assume a common $K$ just for the simplicity.
}
Note that we use the same discretization scheme for all components of $\varphi_n$. 
Here eq.~\eqref{eq:16} is also used, and $K$ is the number of discrete points. 
It is found that $f_{\mu}\left(\varphi_{n}, \varphi_{n+\hat{\mu}}\right) $  is a matrix
whose indices are $\varphi_{n}$ and $\varphi_{n+\hat{\mu}}$ which take the $K^N$ discrete numbers in $S_K^N$.
In this way, we now consider $f_\mu$ as a matrix, and this fact provides a benefit for a numerical treatment;
that is, one can use linear-algebra techniques instead of the functional analysis.
The indices of the tensor will be naturally derived from this matrix structure of $f_\mu$ 
as will be seen in section~\ref{sec:decomposition}.

%
%

\subsubsection{Construction of tensor}
\label{sec:decomposition}

In order to derive the tensor network structure from eq.~\eqref{eq:26},
one needs to separate $\varphi_n$ and $\varphi_{n+\hat{\mu}}$ in $f_\mu$.
If this separation works, the original field $\varphi_n$ can be traced out at each $n$.

Since $f_\mu$ is a symmetric matrix with complex entries in general,
which is found in the previous sections~\ref{sec:symmetric} and~\ref{sec:discretization},
we carry out the Takagi factorization: for $\varphi,\varphi' \in S_K^N$,
\begin{align}
  \label{eq:18}
  &f_{1}\left(\varphi, \varphi' \right) =\sum_{w=1}^{K^N} U_{\varphi w} \sigma_{w} U^{\mathrm{T}}_{w \varphi'}, 
  \\
  \label{eq:18_2}
  & f_{2}\left(\varphi, \varphi' \right) =\sum_{s=1}^{K^N} V_{\varphi s} \rho_{s} V^{\mathrm{T}}_{s \varphi'},
\end{align}
where $U$ and $V$ are unitary matrices, $U^{\mathrm{T}}$ and $V^{\mathrm{T}}$ are the transposes of $U$ and $V$, respectively, 
and $\sigma_{w}$ and $\rho_{s}$ are non-negative.
Note that this factorization depends on the discretization scheme,
which determines the set $S_{K}$.
Instead of the Takagi factorization, we can also use the SVD as seen in the next section.

We thus find that eq.~\eqref{eq:26} is written as 
\begin{align}
  \label{eq:TNR_B_D}
  Z_{\rm B}\left(K\right) = \sum_{\left\{ w, s\right\}} \prod_{n \in \Gamma} {T_{\rm B}}\left(K\right)_{w_n s_n w_{n-\hat 1}  s_{n-\hat 2}},
\end{align}
where 
\begin{align}
  \label{eq:boson_tensor}
  {T_{\rm B}}\left(K\right)_{ijkl} =
  \left(\frac{1}{\sqrt{2\pi}}\right)^{N} 
  \sqrt{ \sigma_i   \rho_j  \sigma_k \rho_l }
  {\sum_{\varphi  \in S_{K}^N}}^{\!\!\left(\mathrm{disc.}\right)}
  U_{\varphi i} V_{\varphi j}
  U_{ \varphi k } V_{\varphi l }
\end{align}
for all indices.
One can verify eq.~\eqref{eq:TNR_B_D} from eq.~\eqref{eq:26}
by applying the factorization in eqs.~\eqref{eq:18} and~\eqref{eq:18_2} 
to $f_\mu\left(\varphi_{n}, \varphi_{n+\hat{\mu}}\right)$ for $\mu=1,2$ with the local indices $w_n,s_n$.
Then the index $w_n$ ($s_{n}$) can be interpreted as a variable defined on the link 
which connects $n$ and $n+\hat 1$ ($n+\hat 2$),
so eq.~\eqref{eq:TNR_B_D} forms a tensor network on the two-dimensional lattice $\Gamma$
as with the case of the fermion partition function in eq.~\eqref{eq:TNR_F}.
Here one finds the correspondence between the tensor indices and the hopping structure of the lattice action as in the fermion part.
From this one can see that the tensor network structure is originated from the kinetic terms for both fermions and bosons.

We expect that, in the large $K$ limit, $Z_{\rm B}\left(K\right)$ converges to $Z_{\rm B}$ with an exact tensor network representation
\begin{align}
  \label{eq:TNR_B}
  Z_{\rm B} 
  =  \sum_{\left\{w,s\right\}} \prod_{n \in \Gamma} {T_{\rm B}}_{w_n s_n w_{n-\hat 1}  s_{n-\hat 2}}
\end{align}
if we can find a proper discretization scheme so that $T_{\rm B}\left(K\right)$ converges
to $T_{\rm B}$ in $K \rightarrow \infty$.
In practice one has to confirm that $Z_{\mathrm{B}}\left(K\right)$ converges to $Z_{\mathrm{B}}$ with increasing $K$ in the choice of a discretization scheme.
We will see this point in section~\ref{sec:Results_Wilsonboson}.

%
%

\subsubsection{Miscellaneous remarks}
\label{sec:remarks}

We give some miscellaneous remarks which may be important for future applications 
and deeper understanding of the symmetry of the tensor network.

The tensor network representation in the form of eq.~\eqref{eq:TNR_B}, 
which gives the boson partition function in eq.~\eqref{eq:ZB}, is not uniquely determined. 
Let $F$ and $G$ be regular $K \times K$ matrices.
We then find that eq.~\eqref{eq:TNR_B} also holds for another uniform tensor $\tilde T_{\rm B}$ given by
\begin{align}
  \tilde {T_{{\rm B}}}_{w s w' s'}= 
  {T_{\rm B}}_{i j k l}
  F_{iw}
  G_{js}  
  F^{-1}_{k w'} 
  G^{-1}_{l s'} 
\end{align}
for all indices. 
Furthermore, by using $F_n$ and $G_n$ that are regular matrices satisfying the periodic boundary conditions
on the two-dimensional lattice $\Gamma$ and transforming 
${T_{\rm B}}_{w_n s_n w_{n-\hat 1}  s_{n-\hat 2}}$ by $F_n$, $G_n$, $F^{-1}_{n-\hat 1}$, and $G^{-1}_{n-\hat 2}$,
$Z_{\rm B}$ can also be written in terms of the non-uniform tensors.
This means that the tensor network representation of the partition function is invariant under the gauge transformations for tensors.

The expression of eq.~\eqref{eq:TNR_B_D} is rather general in the sense that we can always find it 
for two-dimensional PT-invariant theories with the real scalars. 
It is very easy to generalize this result to more complicated cases, non PT-invariant actions 
which, for example, have only one of $\phi^2_{n+\hat \mu} \phi_n$ or $\phi^2_{n-\hat \mu} \phi_n$ terms
or the theories with the complex scalars. For those theories, although $f_\mu$ is not symmetric in general, 
we can use the SVD instead of the Takagi factorization.
Then, $U^{\mathrm{T}}$ and $V^{\mathrm{T}}$ in eqs.~\eqref{eq:18} and~\eqref{eq:18_2} are replaced by other unitary matrices,
and we can express the partition function by a similar construction of the tensor to eq.~\eqref{eq:boson_tensor},
where the second $U$ and the second $V$ are replaced with the other ones.
An extension to the higher-dimensional theories is also straightforward.

We have much simpler expressions for the cases of $S_{\rm B,naive}$ given in eq.~\eqref{eq:SBnaive}
because the auxiliary fields are not needed ($N=1$) and $L_\mu$ is isotropic and given by a single $L$:
\begin{align}
  L\left(\phi,\phi'\right)  = \frac{1}{2} \left(\phi'-\phi\right)^2
  + \frac{1}{8} W^\prime\left(\phi\right)^2  + \frac{1}{8} W^\prime\left(\phi'\right)^2.
\end{align}
Equation~\eqref{eq:boson_tensor} then becomes
\begin{align}
  \label{eq:boson_tensor2}
  {T_{\rm B, naive}}\left(K\right)_{ijkl} =
  \frac{1}{\sqrt{2\pi}}
  \sqrt{ \sigma_i \sigma_j  \sigma_k \sigma_l }
  {\sum_{\phi  \in S_{K}}}^{\!\!\left(\mathrm{disc.}\right)}
  U_{\phi i} U_{\phi j}  
  U_{ \phi k } U_{\phi l }
\end{align}
because $f_\mu  \equiv f =e^{-L}$ for $\mu=1,2$ and because $V=U$ and $\rho_i=\sigma_i$ in eqs.~\eqref{eq:18} and~\eqref{eq:18_2}. 
In this case, instead of the Takagi factorization, we can use the SVD:
\begin{align}
  \label{eq:svd_for_naiveSB}
  f\left(\phi, \phi' \right) =\sum_{w=1}^K O_{\phi w} \sigma_{w} P^{\mathrm{T}}_{w \phi'}, 
\end{align}
where $O$ and $P$ are real symmetric matrices.
Then we have 
\begin{align}
  {T_{\rm B, naive}}\left(K\right)_{ijkl} =
  \frac{1}{\sqrt{2\pi}}
  \sqrt{ \sigma_{i}   \sigma_{j}  \sigma_{k}   \sigma_{l}}
  {\sum_{\phi \in S_{K}}}^{\!\!\left(\mathrm{disc.}\right)}  
  O_{\phi i} P_{\phi j}  
  O_{ \phi k} P_{\phi l}.
\end{align}

%
%

\subsection{Total tensor network}

We have seen that the fermion and the boson partition functions can be expressed as the tensor networks 
in the previous two sections. 
By combining these results, we can also express the total partition function as a tensor network.

Before presenting the total tensor, let us introduce combined indices $X_n, T_n$.
$X_{n}$ is defined as $X_n = \left(u_n,v_n,w_n\right)$, where ($u_n,v_n$) and $w_n$ are indices of the fermion and the boson tensors, respectively. 
$T_n$ is also defined as $T_{n} = \left(p_n,q_n,s_n\right)$,
and the dimension of $X_n$ and $T_n$ is $2\times2\times K^N$.

The total tensor is made by replacing $e^{-S_{\rm B}\left(\phi\right)}$ in eq.~\eqref{eq:ZB} with  $e^{-S_{\rm B}\left(\phi\right)}Z_{\rm F}\left(\phi\right)$
and repeating the same procedure for making the tensor network representation of the boson partition function. 
Additional contributions by $Z_{\rm F}$ do not give any complexity. 
We find that the total tensor network representation is given by the boson one in eq.~\eqref{eq:TNR_B_D} multiplied by  $Z_{\rm F}\left(\phi\right)$ from the right: 
\begin{align}
  \label{eq:19}
  Z\left(K\right) = \sum_{\left\{X,T\right\}} &\prod_{n \in \Gamma} {\mathcal{T}\left(K\right)}_{X_{n} T_{n} X_{n-\hat{1}} T_{n-\hat{2}}}
                                                \nonumber \\
                                              & \cdot 
                                                \int {\cal D} \Xi_{uvpq} 
                                                \prod_{n \in \Gamma} 
                                                \left(\bar \xi_{n+\hat 1}\xi_n\right)^{u_n}  \left(\bar \chi_{n+\hat 1}\chi_n\right)^{v_n}  \left(\bar \eta_{n+\hat 2}\eta_n\right)^{p_n}  \left(\bar \zeta_{n+\hat 2}\zeta_n\right)^{q_n}
\end{align}
with
\begin{align}
  {\mathcal{T}\left(K\right)}_{X T X' T'} =
  \left(\frac{1}{\sqrt{2\pi}}\right)^{N}
  \sqrt{\sigma_{w}   \rho_{s}  \sigma_{w'}   \rho_{s'}}
  {\sum_{\varphi \in S_{K}^N}}^{\!\!\left(\mathrm{disc.}\right)}  
  U_{\varphi w} 
  V_{\varphi s}  
  U_{ \varphi w'} 
  V_{\varphi s'} 
  {T_{\rm F}\left(\phi\right)}_{u v p q u' v' p' q' },
  \label{eq:total_T}
\end{align}
where $U$, $V$, $\sigma_w$, and $\rho_s$ are given 
by eqs.~\eqref{eq:18} and~\eqref{eq:18_2}.
The measure ${\cal D} \Xi_{uvpq} $ is given in eq.~\eqref{eq:measures}, 
$\xi, {\bar \xi}, \chi, {\bar \chi}, \eta, {\bar \eta}, \zeta, {\bar \zeta}$ are one-component Grassmann numbers,
and $T_{\rm F}$ is the tensor for the fermion part defined in eq.~\eqref{eq:fermion_tensor}. 
Note that $T_{\rm F}\left(\phi_n\right)$ includes only $\phi_n$
which is a component of $\varphi_n$.
The total tensor ${\mathcal{T}\left(K\right)}_{X T X' T'}$ is uniformly defined on the lattice.

Now the original partition function $Z$ is expressed as a tensor network $Z\left(K\right)$.
We have built it for a general superpotential by focusing on the hopping structure of the lattice action.
Introduction of local interaction terms does not change our formulation but rather elements of tensor.
Moreover, the same structure of the tensor network leads to the same order of computational complexity.\footnote{
  If the superpotential contains hopping terms, the hopping structure of the lattice action changes and one has to slightly modify the derivation of the tensor network representation.
}
We will numerically verify that $Z\left(K\right)$ indeed converges to $Z$ by using the TRG as a coarse-graining scheme for the tensor network in the next section.

%
%
%
%

\section{Numerical test in free theory}
\label{sec:Results}

The partition function of the lattice $\mathcal{N}=1$ Wess--Zumino model has been expressed as a tensor network in 
eq.~\eqref{eq:19}. In this section, we test the expressions in the free theory given by eq.~\eqref{eq:free_P}
varying the mass for three lattice sizes $V=2 \times 2$, $8 \times8$, $32\times 32$ 
with the periodic boundary conditions. 
Numerical tests in the free theory are effective to study whether the tensor is correctly given by our new formulation
because the tensor network structure is derived from the hopping terms in the action, i.e. the kinetic terms.
The computation is performed with the value of the Wilson parameter $r=1/\sqrt{2}$ to reduce the computational cost
because the auxiliary field $G$ is decoupled as seen in section~\ref{sec:auxiliary_field}.

%
%

\subsection{Some details}
\label{sec:Some_details}

In sections~\ref{sec:Results_Majorana} and~\ref{sec:Results_Wilsonboson}, 
we compute $Z_{\rm F}$ and $Z_{\rm B}$ individually using the (Grassmann) TRG
since they are independent with each other in the free theory. 
In section~\ref{sec:Results_Wilsonboson},  the Witten index given by the total partition function $Z$ is computed
by the Grassmann TRG.
Since the free theory is exactly solvable, we can compare an obtained result $X_{\mathrm{TRG}}$ 
with the exact solution $X_{\mathrm{exact}}$ by computing 
\begin{align}
  \label{eq:delta}
  \delta(X)
  = \left| \frac{X_{\mathrm{exact}} - X_{\mathrm{TRG}}}{X_{\mathrm{exact}}}\right|.
\end{align}

In what follows, we briefly describe the TRG while introducing $D_{\mathrm{cut}}$ which defines the truncated dimension of tensors.
The SVD allows us to express a tensor $T_{ijkl}$ ($i,j,k,l=1,2,\cdots,N$) of which the tensor network representation of a partition function $\mathcal{Z}$ is made 
as $T_{ijkl}= \sum_{I=1}^{N^2} S_{ijI} \sigma_{I}  (V^\dag)_{Ikl}$, where 
$S$ and $V$ are unitary matrices and $\sigma_I$ is the singular value of $T_{ijkl}$.
We assume that the singular values are sorted in descending order: $\sigma_1 \ge \sigma_2 \ge \sigma_3 \ge \cdots \sigma_{N^2} \ge 0$.\footnote{
  Strictly speaking, $S$ and $V$ are matrices with respect to the row specified by $i,j$ and the column $I$, 
  and $\sigma_I$ is the singular values of the matrix $T_{ijkl}$ with the row $i,j$ 
  and the column $k,l$. In addition, $S$ and $V$ are taken to be real symmetric ones 
  when $T_{ijkl}\in \mathbb{R}$ for all $i,j,k,l$.
}
In the TRG, 
$T_{ijkl}$ is approximately decomposed:
\begin{align}
  T_{ijkl} \approx \sum_{I=1}^{D_{\rm cut}} S_{ijI} \sigma_{I}  (V^\dag)_{Ikl} ,  
  \label{eq:cg_tensor}
\end{align}
where $D_{\rm cut}$, which is fixed throughout a computation, is used to truncate the dimension of the tensor indices if it is smaller than $N^2$.
If not so, the summation in eq.~\eqref{eq:cg_tensor} is done up to $N^2$ without the truncation.
A similar decomposition can be done with a different combination of the indices:
\begin{align}
  T_{ijkl} \approx \sum_{I=1}^{D_{\rm cut}} S_{liI}^\prime \sigma_{I}^\prime  (V^{\prime\dag})_{Ijk} .
  \label{eq:cg_tensor_2}
\end{align}
The coarse-grained tensor $T^{\rm new}_{IJKL}$ with $I,J,K,L=1,\ldots,{\rm min}\{D_{\rm cut},N^2\}$ 
is then given by contracting the rank-three tensors
$\sqrt{\sigma} S, \sqrt{\sigma^\prime} S^\prime, \sqrt{\sigma} V, \sqrt{\sigma^\prime} V^\prime$
and forms a network again as with $T_{ijkl}$. 
We can compute the partition function $\mathcal{Z}$ by repeating this procedure.
Since the number of tensors decreases through the coarse-graining,
$\mathcal{Z}$ is finally given by a single tensor for which the indices are contracted: $\mathcal{Z}=\sum_{I,J=1}^{D_{\rm cut}} T^{\rm new}_{IJIJ}$. 
More details are shown in ref.~\cite{Gu:2009dr}, and appendix~\ref{sec:Coarse-graining_GTRG} is given for the Grassmann cases.

We employ the Gauss--Hermite quadrature~\eqref{eq:GH-quardrature} 
to discretize the integrals of $\phi$ and $H$ in~\eqref{eq:16}:
\begin{align}
  \label{ZB_used}
  Z_{\mathrm{B}}\left(K\right)
  = \prod_{n \in \Gamma} \left(\frac{1}{\sqrt{2\pi}}\right)^{2}
  {\sum_{\phi_{n} \in S_{K}}}^{\!\!\!\!\left(\mathrm{GH}\right)}
  {\sum_{H_{n} \in S_{K}}}^{\!\!\!\!\left(\mathrm{GH}\right)}
  \prod_{\mu=1}^{2} f_{\mu}\left(\phi_n, H_n, \phi_{n+\hat{\mu}}, H_{n+\hat{\mu}}\right),
\end{align}
where
\begin{align}
  \label{fmu_used}
  f_\mu\left(\varphi_n, \varphi_{m}\right)
  = \exp \biggl\{&-\frac{1}{2}\left(1+\frac{m}{\sqrt{2}}\right) \left(\phi_n-\phi_m\right)^2   
                   - \frac{m^{2}}{8} \left(\phi_n^2 + \phi_m^2\right) \nonumber \\
                 &- \frac{1}{8} \left(H_{n}^{2} + H_{m}^{2} \right)
                   -\frac{\left(-1\right)^{\delta_{\mu2}}}{2\sqrt{2}} \left(H_{n}-H_m\right) \left(\phi_n -\phi_m\right)\biggr\}
\end{align}
for eq.~\eqref{eq:free_P} and $r = 1/\sqrt{2}$. 
The two-dimensional variable
$\varphi_n= \left(\varphi_{n,1},\varphi_{n,2}\right) = \left(\phi_n/\sqrt{2\pi}, H_n/\sqrt{2\pi}\right)$ 
is again used for the notational simplicity.
$S_K$ is a set of the roots of the $K$-th Hermite polynomial.
We use the SVD to decompose $f_\mu$, which are $K^2 \times K^2$ real symmetric matrices, as
\begin{align}
  \label{f1_svd_used}
  &f_1(\varphi,\varphi') = \sum_{w=1}^{K^2}  O_{\varphi w} \sigma_{w} P_{\varphi' w}, \\
  \label{f2_svd_used}
  &f_2(\varphi,\varphi') = \sum_{s=1}^{K^2}  S_{\varphi s} \rho_{s} T_{\varphi' s},
\end{align}
where $\sigma_1 \ge\sigma_2 \ge \ldots \ge \sigma_{K^2}$ and 
$\rho_1 \ge \rho_2 \ge \ldots \ge \rho_{K^2}$. 
For reducing the memory usage and the computational cost,
we initially approximate the tensor network representation of eq.~\eqref{ZB_used} by $D_{\mathrm{init}} \le K^{2}$:
\begin{align}
  \label{ZB_K_used}
  Z_{\rm B}\left(K\right) \approx \prod_{n \in \Gamma} \sum_{w_n=1}^{D_{\mathrm{init}}}  \sum_{s_n=1}^{D_{\mathrm{init}}} {T_{\rm B}}\left(K\right)_{w_n s_n w_{n-\hat 1} s_{n-\hat 2}},
\end{align}
where 
\begin{align}
  \label{boson_tensor_used}
  {T_{\rm B}}\left(K\right)_{ijkl} =
  \frac{1}{2\pi} 
  \sqrt{\sigma_i \rho_j \sigma_k \rho_l}
  {\sum_{\phi_{n} \in S_{K}}}^{\!\!\!\!\left(\mathrm{GH}\right)}
  {\sum_{H_{n} \in S_{K}}}^{\!\!\!\!\left(\mathrm{GH}\right)}
  O_{\varphi i} S_{\varphi j}  
  P_{ \varphi k } T_{\varphi l}.
\end{align}
Note that $D_{\mathrm{init}}$ defines the bond dimension of the initial tensor. 
We will simply take $D_{\mathrm{init}}=D_{\rm cut}$ for evaluating $Z_{\rm B}$ in section~\ref{sec:Results_Wilsonboson}  
and $D_{\mathrm{init}}=D_{\rm cut}/2$ for the Witten index  in section~\ref{sec:Results_freeWZ}  
because the bond dimension does not change after the coarse-graining steps under these choices.

Here we mention the computational costs for the coarse-graining of tensor networks and for the construction of tensors.
Both of them are mainly consists of the SVD and the contraction of tensor indices.
Since the cost of the numerical SVD for square matrices is proportional to the third power of the matrix dimension,
the computational effort required for the numerical decomposition described in eq.~\eqref{eq:cg_tensor} and in eqs.~\eqref{f1_svd_used} and~\eqref{f2_svd_used} are in proportion to $N^{6}$ and $K^{6}$, respectively.
A contraction of tensor indices is expressed as a summation of them, so the cost of the contraction depends on the number of the tensor indices.
Then it is proportional to ${D_{\mathrm{cut}}}^{6}$ when contracting the rank-three tensors described around eqs.~\eqref{eq:cg_tensor} and~\eqref{eq:cg_tensor_2},
and is proportional to $K^{2} \times {D_{\mathrm{init}}}^{4}$ when building the tensor in eq.~\eqref{boson_tensor_used}.
For the coarse-graining step, one can find that the volume-dependence of the cost is milder than $D_{\mathrm{cut}}$-dependence as follows.
Since the TRG is a coarse-graining of space-time, one can reach a large space-time volume by simply iterating the same local blocking procedures.
More directly, the computational cost of the TRG is proportional to the logarithm of the space-time volume, i.e. the number of iterations.
Summarizing the above, the computational cost for the coarse-graining of tensor networks is proportional to ${D_{\mathrm{cut}}}^{6} \times \ln V$,
and that for the construction of tensors is proportional to $\max\left\{K^{6}, K^{2}\times {D_{\mathrm{cut}}}^{4}\right\}$,
where $N = D_{\mathrm{init}} = D_{\mathrm{cut}}$ is simply assumed.

%
%

\subsection{Free Majorana--Wilson fermion}
\label{sec:Results_Majorana}

\begin{figure}[htbp]
  \centering
  \includegraphics[width=0.8\hsize]{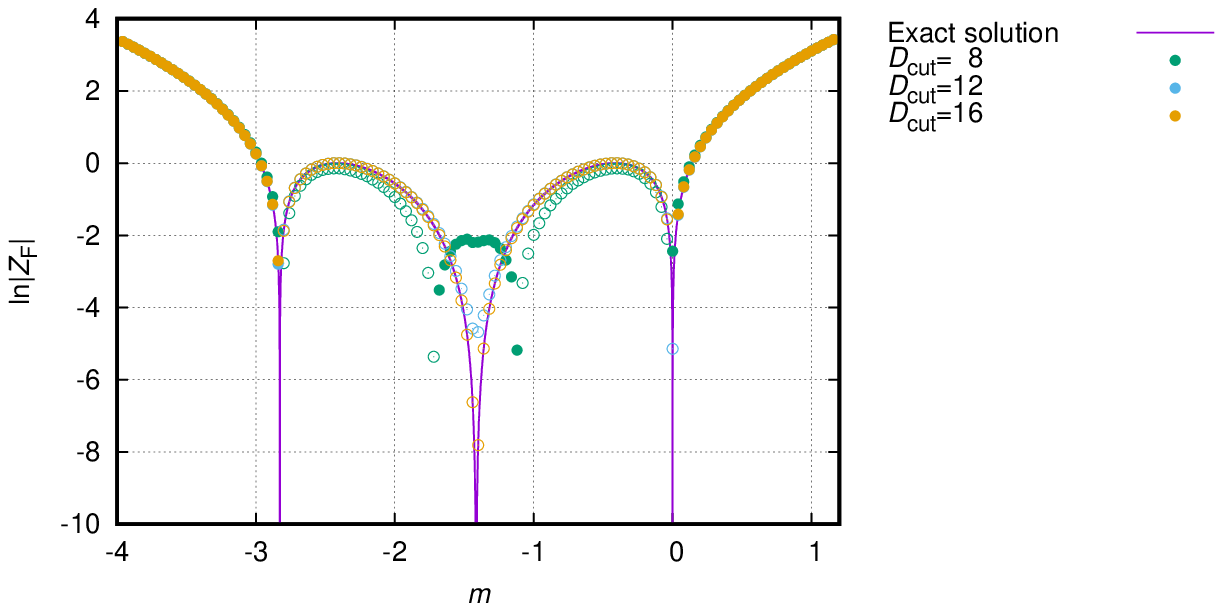}
  \includegraphics[width=0.8\hsize]{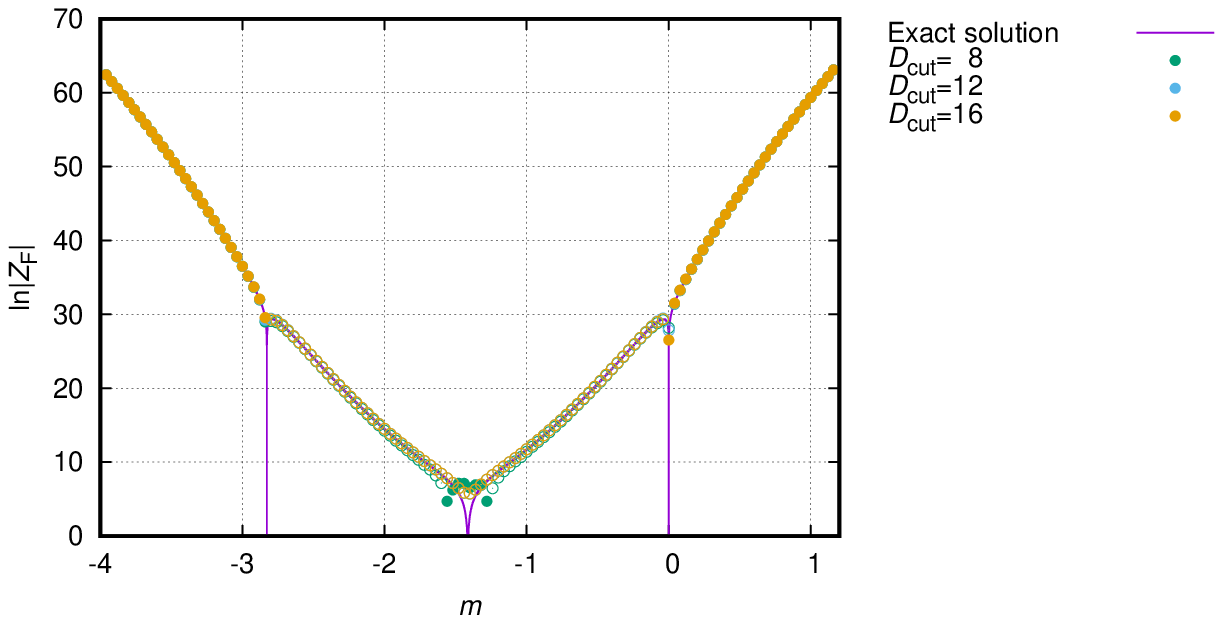}
  \includegraphics[width=0.8\hsize]{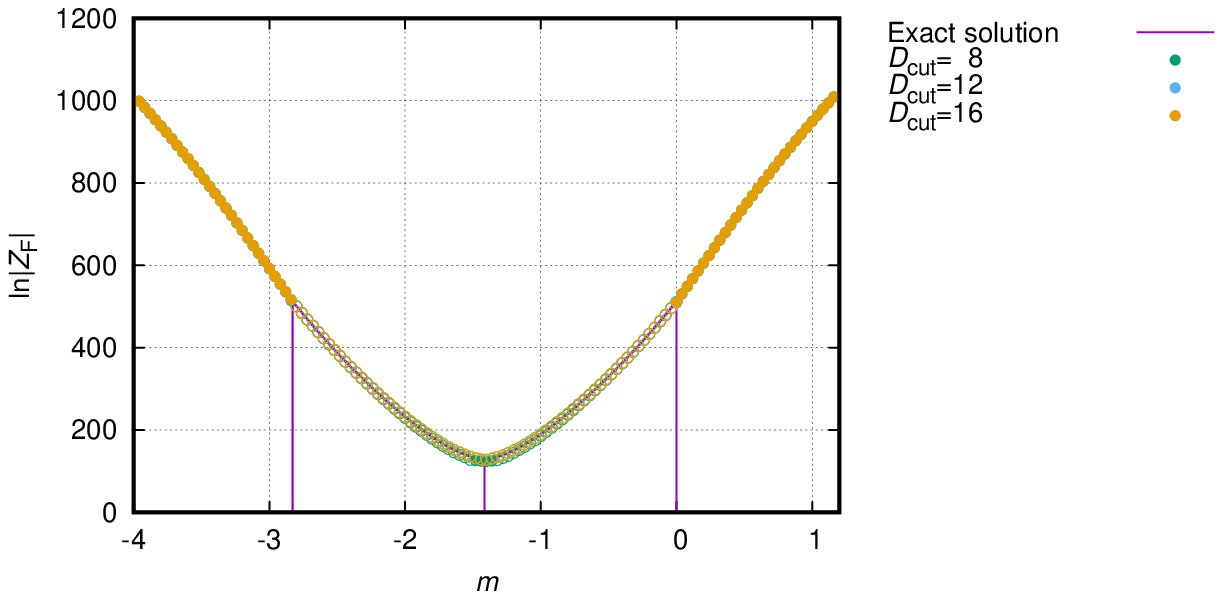}
  \caption{$\ln \left(\left|Z_{\rm F}\right|\right)$ of free Majorana-Wilson fermions with $r=1/\sqrt{2}$ is plotted against $m$ for $V=2\times 2$~(top), $8\times 8$~(center), $32\times 32$~(bottom).
    The solid (open) symbols represent the positive (negative) sign of $Z_{\rm F}$.
    \label{fig:ZF}}
\end{figure}

Figure~\ref{fig:ZF} shows the logarithm of the fermion Pfaffian computed by the Grassmann TRG 
with varying $m$ for $V=2\times 2$~(top), $8\times 8$~(center), $32\times 32$~(bottom).  
The green, blue, and yellow symbols denote the results for three different bond dimensions: $D_{\rm cut}=8$, $12$, $16$,
and the solid and open ones indicate the positive and negative sign of the Pfaffian, respectively.
The purple curves represent the exact solutions given by eq.~\eqref{eq:Z_F_exact}. Three negative peaks 
at $m=0$, $-\sqrt{2}$, $-2\sqrt{2}$ correspond to the fermion zero modes,
and the exact Pfaffian has the negative sign for $-2\sqrt{2}<m<0$ as can be seen in eq.~\eqref{eq:Z_F_exact}.

In the top plot of figure~\ref{fig:ZF}, 
the green symbols ($D_{\rm cut}=8$) around the peak at the center are rather deviated from the exact solution, 
and they even have the opposite sign. 
The deviation becomes smaller as $D_{\rm cut}$ increases, and the yellow symbols ($D_{\rm cut}=16$) have the correct sign and 
agree well with the exact one even near the peak.
The situation is further improved by taking larger volumes  even for the smallest $D_{\rm cut}$, 
and the numerical results fit well with the analytical curve in the center and the bottom figures.

\begin{figure}[htbp]
  \centering
  \includegraphics[width=0.7\hsize]{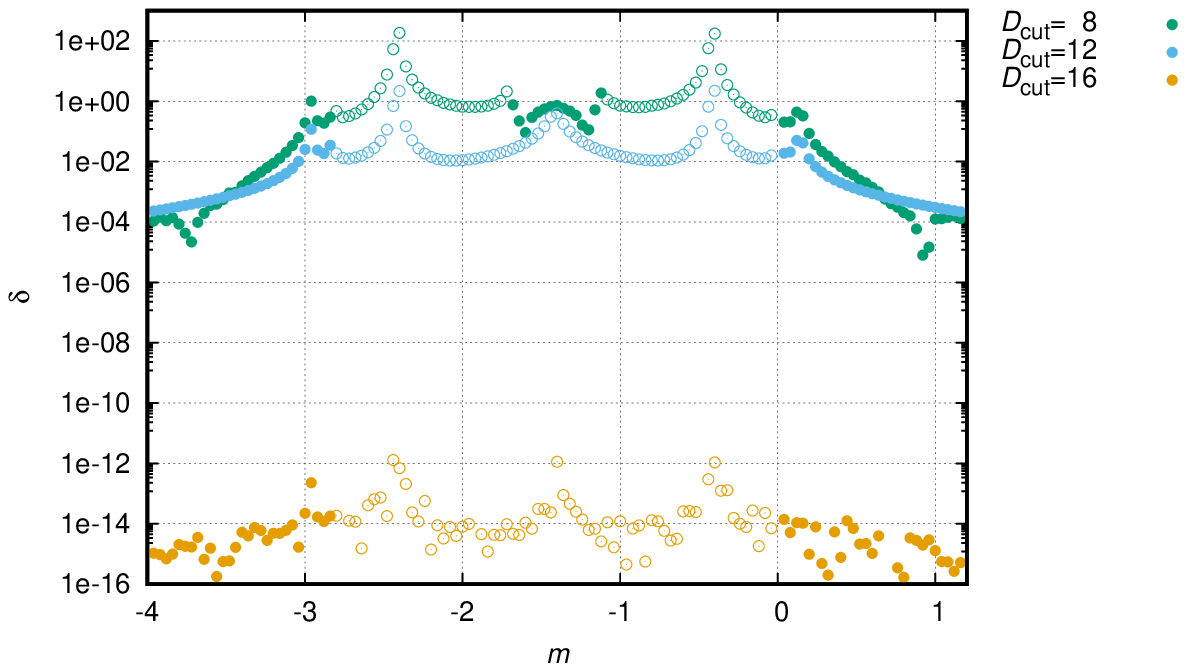}
  \includegraphics[width=0.7\hsize]{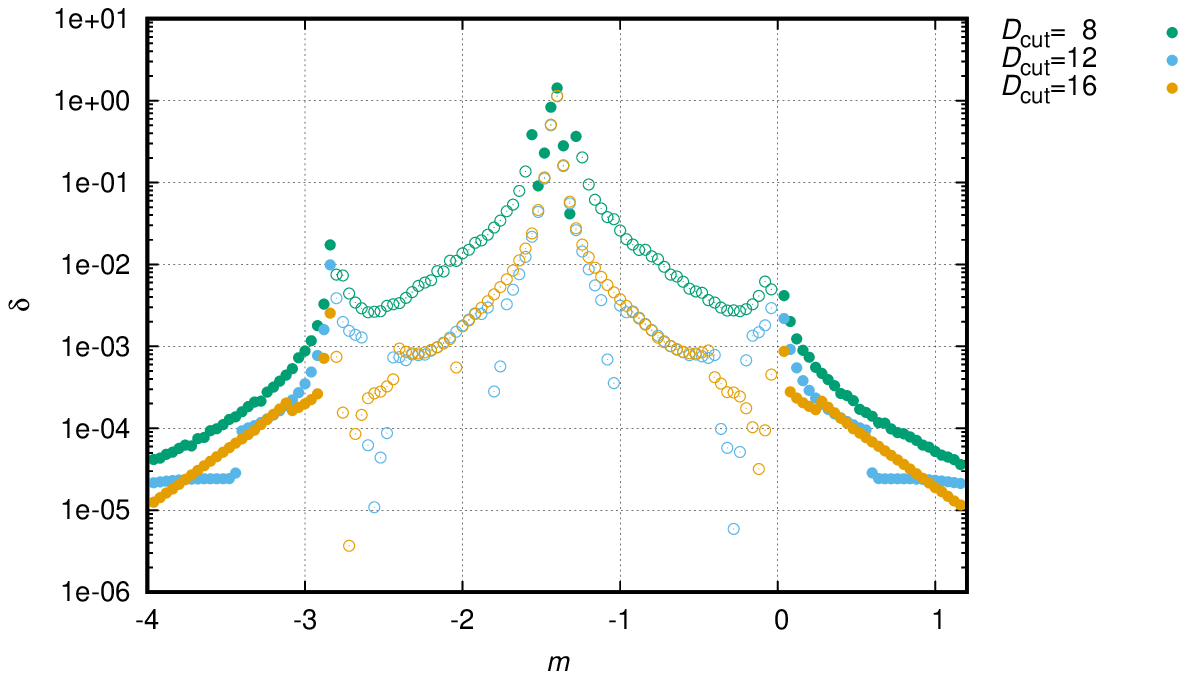}
  \includegraphics[width=0.7\hsize]{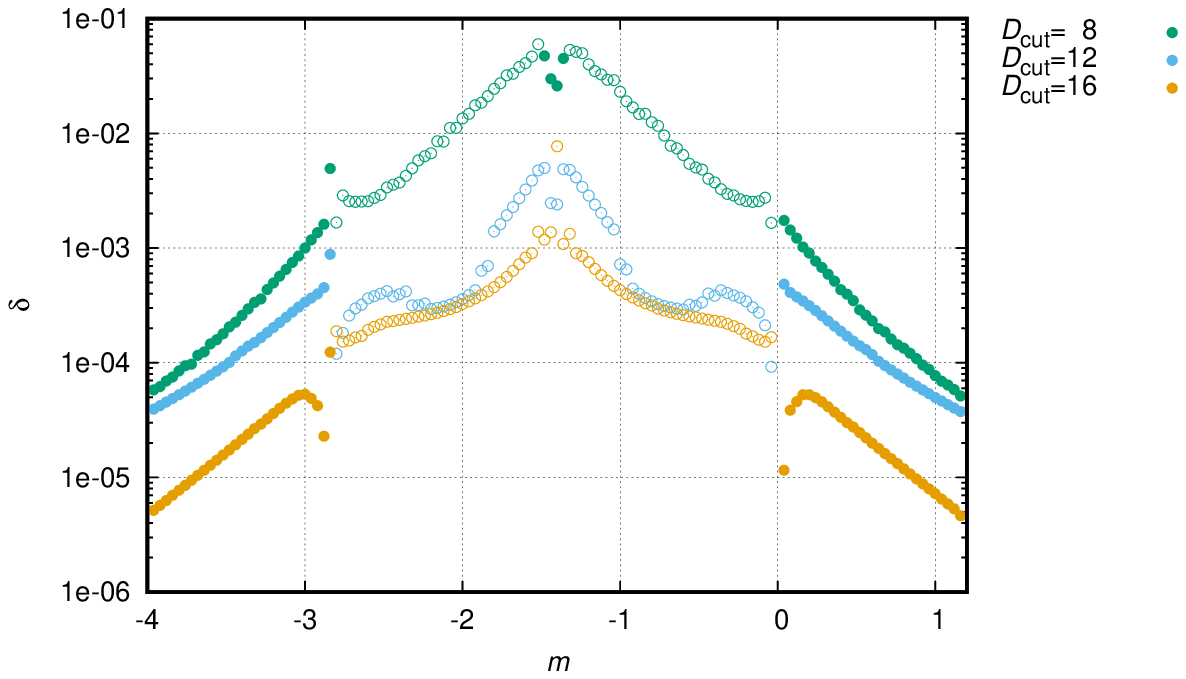}
  \caption{Relative errors of $\ln\left(\left|Z_{\rm F}\right|\right)$ against $m$.  The results are shown for $V=2\times 2$~(top), $8\times 8$~(center), $32\times 32$~(bottom).
    The solid (open) symbols represent the positive (negative) signs of $Z_{\rm F}$, respectively.
    \label{fig:ZF_error}}
\end{figure}

These observation can also be clearly understood in figure~\ref{fig:ZF_error},
which shows the relative errors $\delta\left(\ln \left|Z_{\rm F}\right|\right)$ given by eq.~\eqref{eq:delta}. 
Note that the case for $D_{\rm cut}=16$ on $V=2\times 2$ have extremely small errors. 
This is because the maximal bond dimension of the coarse-grained tensors on $V=2 \times 2$ lattice is less than or equal to $D_{\rm cut}$.
In other words, no truncation occurs in the TRG steps.
This striking feature is only found in the pure fermion case.
In contrast, the discretization error and the truncation error are inevitable in the boson case
since the approximation already enters in deriving the tensor network representation of the boson partition function,
and furthermore the tensor indices are truncated to carry out the numerical evaluation as seen in previous section.
For all volumes used in the computation,  the relative errors almost monotonically decreases
as $D_{\rm cut}$ increases.

Thus we can conclude that the Pfaffian with the correct sign is reproduced 
from the tensor network representation in eq.~\eqref{eq:TNR_F} with eq.~\eqref{eq:fermion_tensor}
using the Grassmann TRG within tiny errors $\mathcal{O}\left(10^{-3}\right)$ 
for physically important parameters, $|m| \ll 1$, and larger volumes.

%
%

\subsection{Free Wilson boson}
\label{sec:Results_Wilsonboson}

The boson partition function is given as a discretized form $Z_{\rm B}\left(K\right)$ 
in eq.~\eqref{ZB_used} by applying the Gauss--Hermite quadrature 
to the integrals of $\phi$ and $H$. 
Then $K$ is the number of the discrete points. 
We prepare the initial tensor network approximately as eq.~\eqref{ZB_K_used} 
and compute it using the TRG for $m > 0$
because the adopted quadrature does not effectively work for $m<0$ (we will see this point later.).
It is, however, sufficient to study the case of $m>0$ 
because the boson action does not depend on the sign of $m$, but on $m^2$, in the continuum theory.

\begin{figure}[htbp]
  \centering
  \includegraphics[width=0.6\hsize]{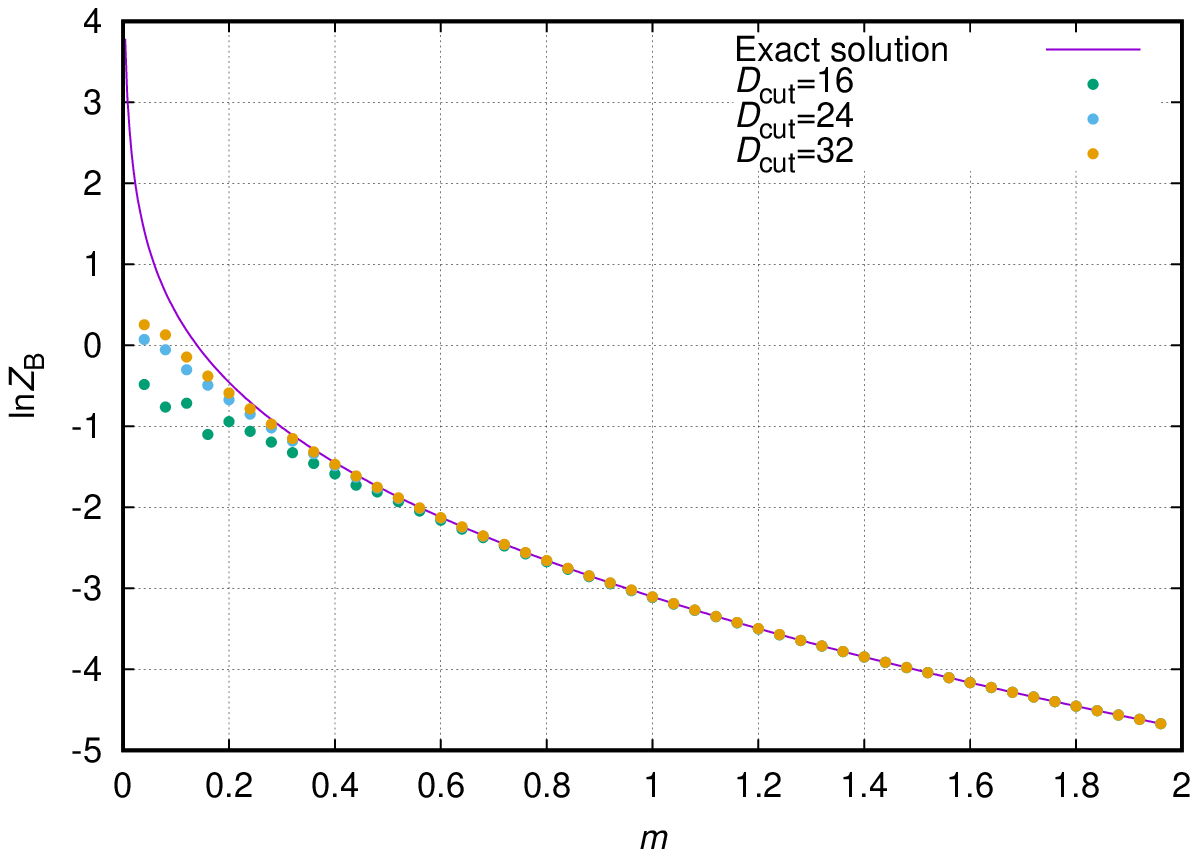}
  \includegraphics[width=0.6\hsize]{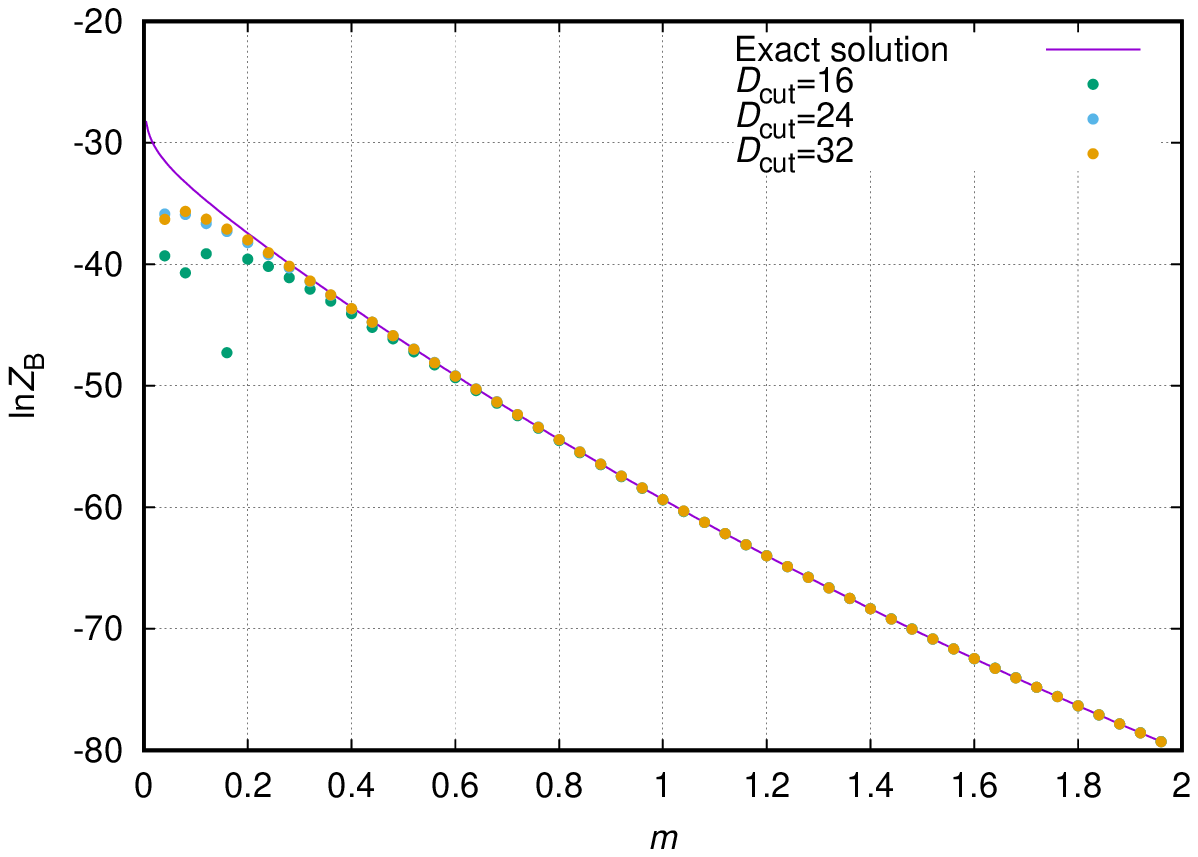}
  \includegraphics[width=0.6\hsize]{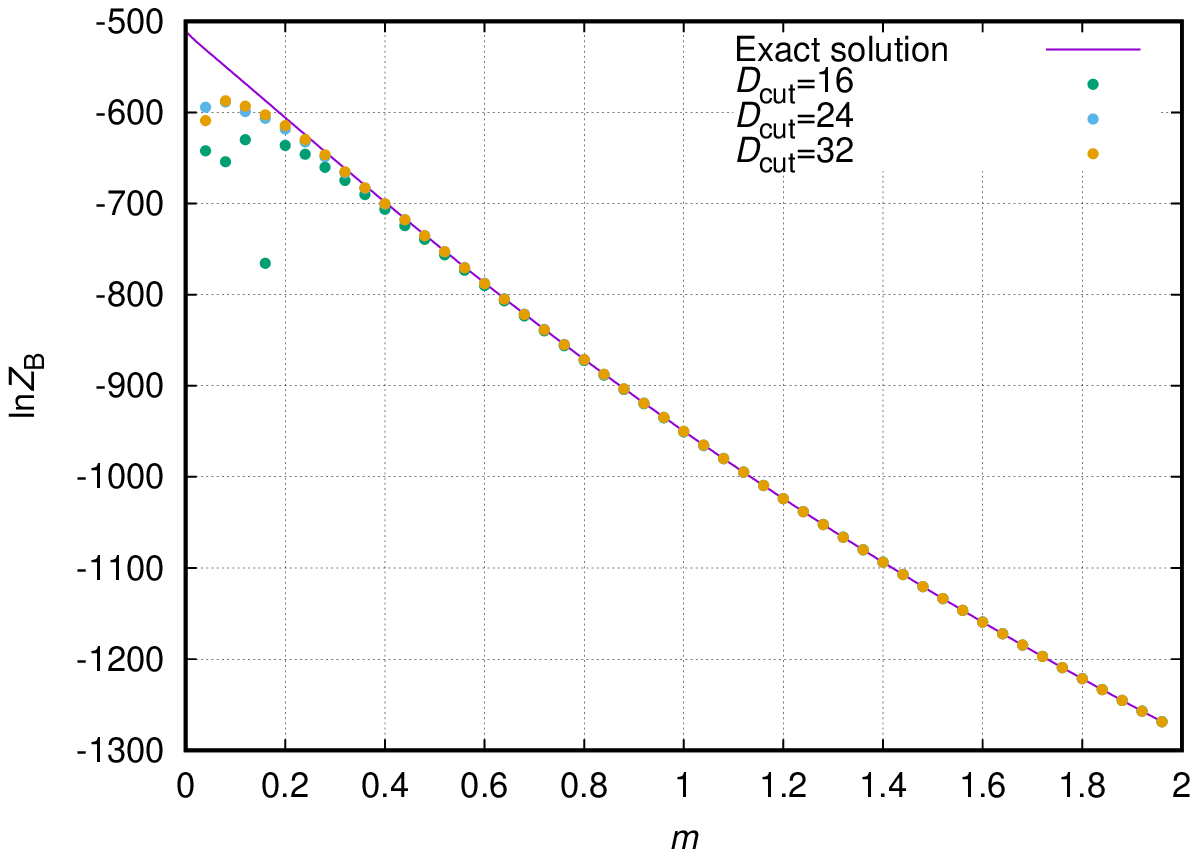}
  \caption{$\ln \left(Z_{\rm B}\left(K\right)\right)$ of free Wilson bosons with $r=1/\sqrt{2}$ against $m$ 
    for $V=2\times 2$~(top), $8\times 8$~(center), $32\times 32$~(bottom).
    The degree of the Hermite polynomial is fixed as $K=64$.
    \label{fig:ZB}}
\end{figure}

\begin{figure}[htbp]
  \centering
  \includegraphics[width=0.6\hsize]{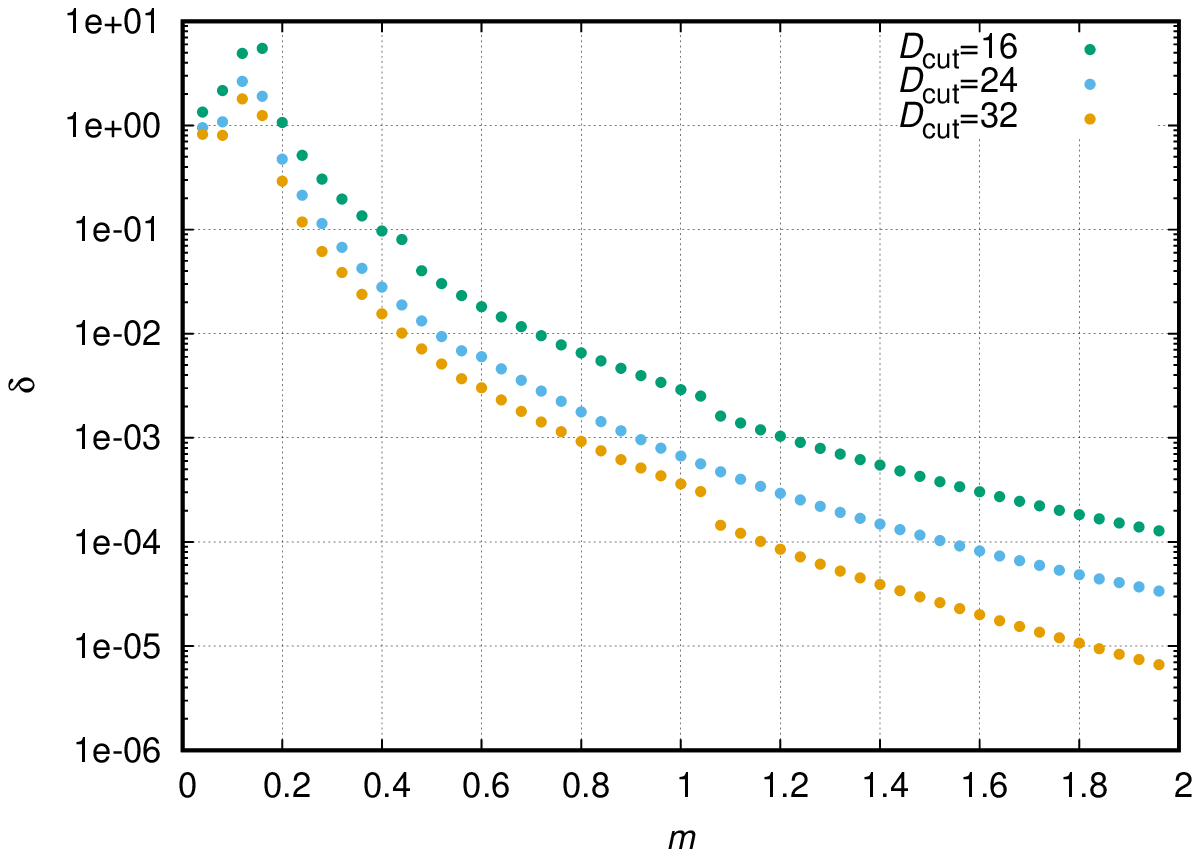}
  \includegraphics[width=0.6\hsize]{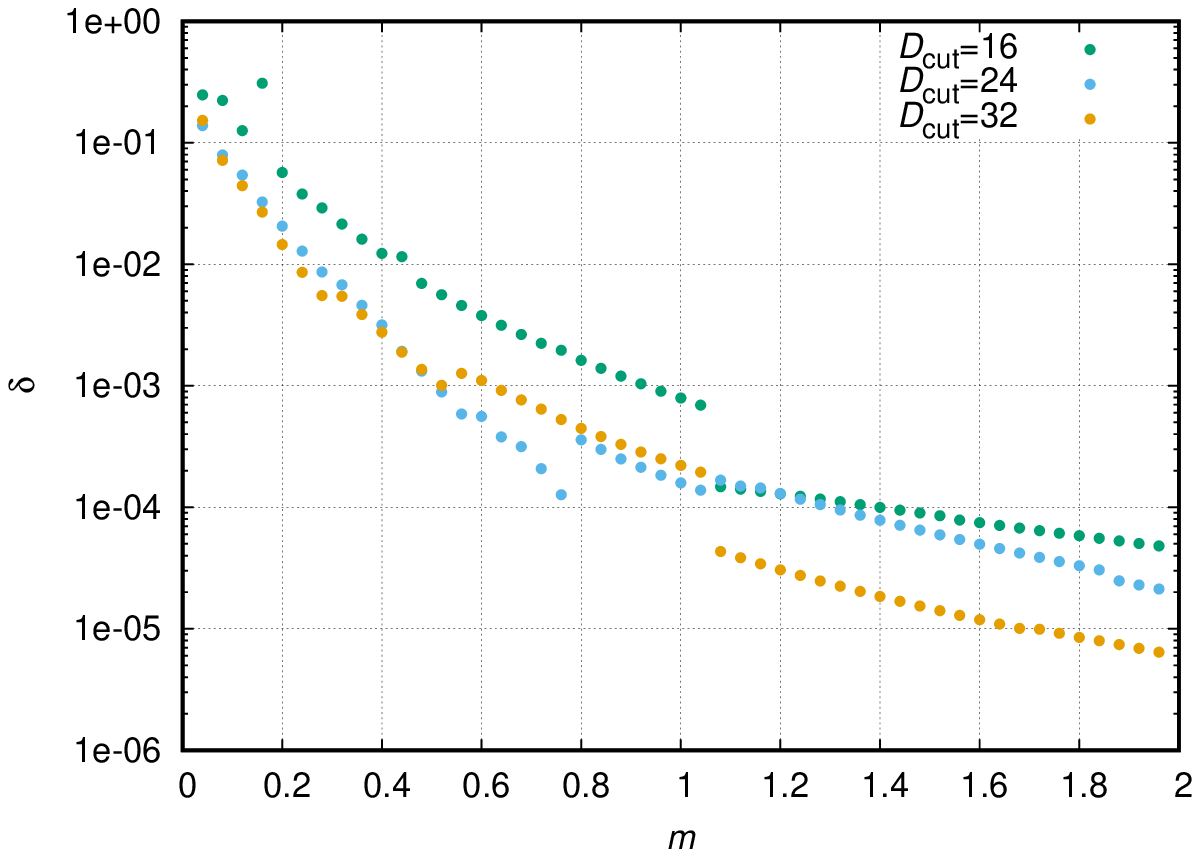}
  \includegraphics[width=0.6\hsize]{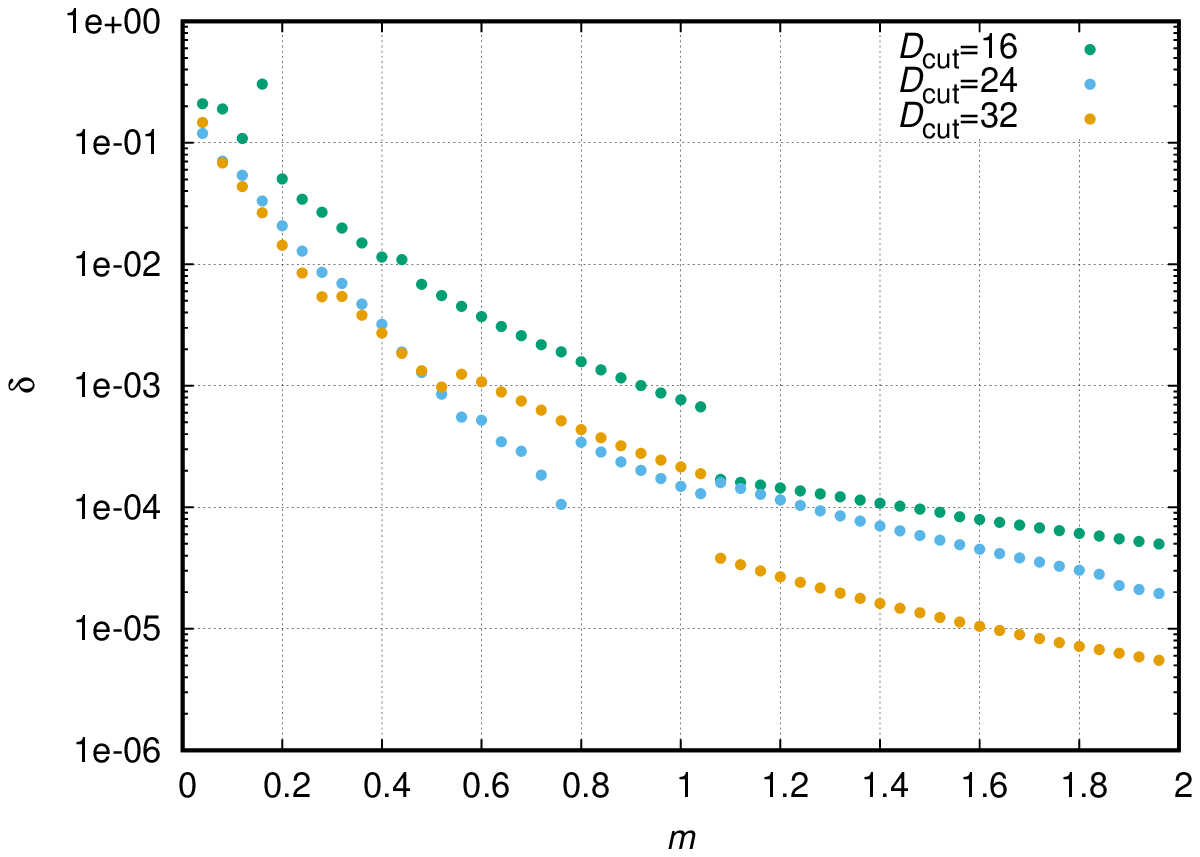}
  \caption{Relative errors of $\ln \left(Z_{\rm B}\left(K\right)\right)$ against $m$ with fixed $K=64$. 
    Top, center, and bottom figures show the results for $V=2 \times 2$, $8\times 8$ and $32 \times 32$, respectively. 
    \label{fig:ZB_error}}
\end{figure}

\begin{figure}[htbp]
  \centering
  \includegraphics[width=0.9\hsize]{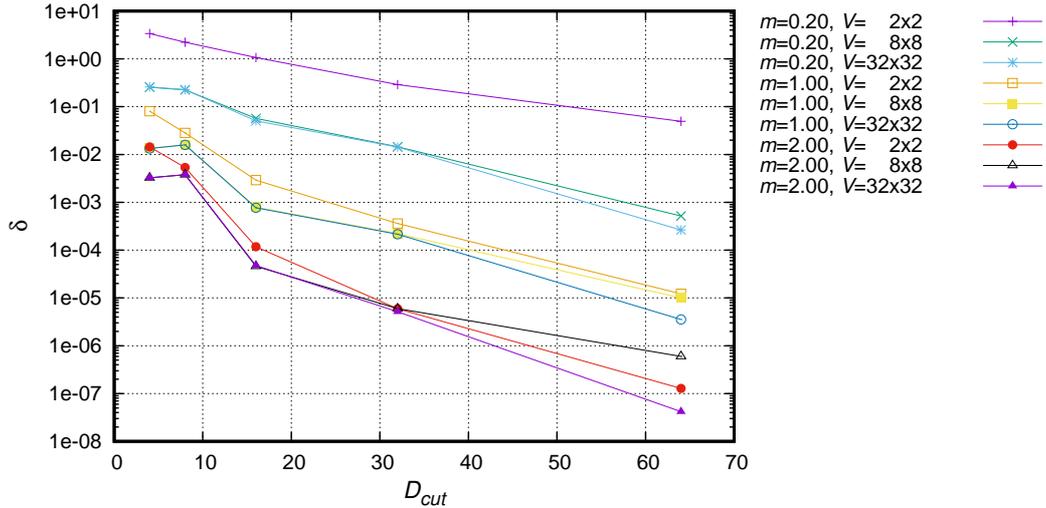}
  \caption{$D_{\rm cut}$-dependence of relative errors of $\ln \left(Z_{\rm B}\left(K\right)\right)$ with $K=64$. }
  \label{fig:ZB_Dcut_dependence}
\end{figure}

Figure~\ref{fig:ZB} shows the logarithm of $Z_{\rm B}\left(K\right)$ with fixed $K=64$, 
and figure~\ref{fig:ZB_error} shows the corresponding relative errors defined by eq.~\eqref{eq:delta}. 
One can see that the TRG results are consistent with the exact ones for large $m$ 
in all of the lattice sizes and $D_{\rm cut}=16$, $24$, $32$. 
Figure~\ref{fig:ZB_Dcut_dependence} shows that the results are systematically improved 
by increasing $D_{\rm cut}$ as one expects.
The exponential improvement may be explained as follows.
Usually the singular values of the tensor are exponentially decaying;
thus from a local point of view the truncation error gets exponentially smaller by increasing $D_{\mathrm{cut}}$.
Since the free energy consists of the local tensors,
it is likely that its error shows such a behavior as well.

\begin{figure}[htbp]
  \centering
  \includegraphics[width=0.6\hsize]{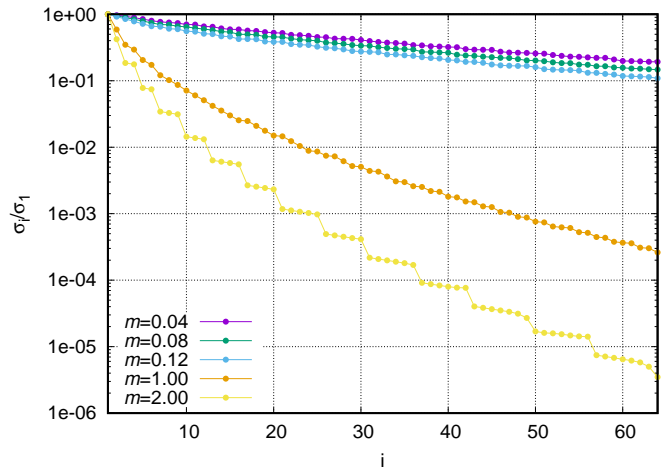}
  \caption{Hierarchy of the singular values of the initial boson tensor for several masses with $K=64$.}
  \label{fig:singular_values_of_initial_tensor}
\end{figure}

The growth of the errors is observed near $m=0$.
Roughly speaking, this is because the massless theory has no damping
factors in $f_\mu$ of eq.~\eqref{fmu_used}.
We can show that $f_\mu$ is expressed as
\begin{align}
  f_\mu\left(\varphi,\varphi'\right)
  = \exp \Biggl\{&- \frac{1}{8}\left(H + \left(-1\right)^{\delta_{\mu 2}}\sqrt{2}\left(\phi - \phi^{\prime}\right)\right)^{2}
                   - \frac{1}{8}\left(H^{\prime} - \left(-1\right)^{\delta_{\mu 2}}\sqrt{2}\left(\phi - \phi^{\prime}\right)\right)^{2} \nonumber \\
                 &- \frac{m^{2}}{16}\left(\phi + \phi^{\prime}\right)^{2}
                   - \frac{m^{2}}{16}\left(1 + \frac{4\sqrt{2}}{m}\right)\left(\phi - \phi^{\prime}\right)^{2} \Biggr\}.
\end{align}
One can see that the damping factors are actually provided for $m>0$ with the damping rate $m^2$ but is not for $-4\sqrt{2} < m < 0$ on the line $\phi=-\phi'$,
so the quadrature does not work for $m<0$.
For $m>0$, we have to take $K$ larger as $m$ decreases so that the quadrature retains effective.
That structure is encoded in the initial tensor in eq.~\eqref{boson_tensor_used} via the matrices $O,P,S,T$
and the singular values $\sigma_w,\rho_s$ in eqs.~\eqref{f1_svd_used} and~\eqref{f2_svd_used}.
The singular values of the initial tensor have unclear hierarchies for small masses as seen in
figure~\ref{fig:singular_values_of_initial_tensor}.
Thus we find that, if $m$ approaches zero from the right, we have to take $K$ and $D_{\rm cut}$ as large as possible to obtain the precise result.\footnote{
  Such a bad behavior could go away once the $\phi^{4}$ interaction term is introduced into the action because it provides the fast damping factor in $f_\mu$.
}

\begin{figure}[htbp]
  \centering
  \includegraphics[width=0.6\hsize]{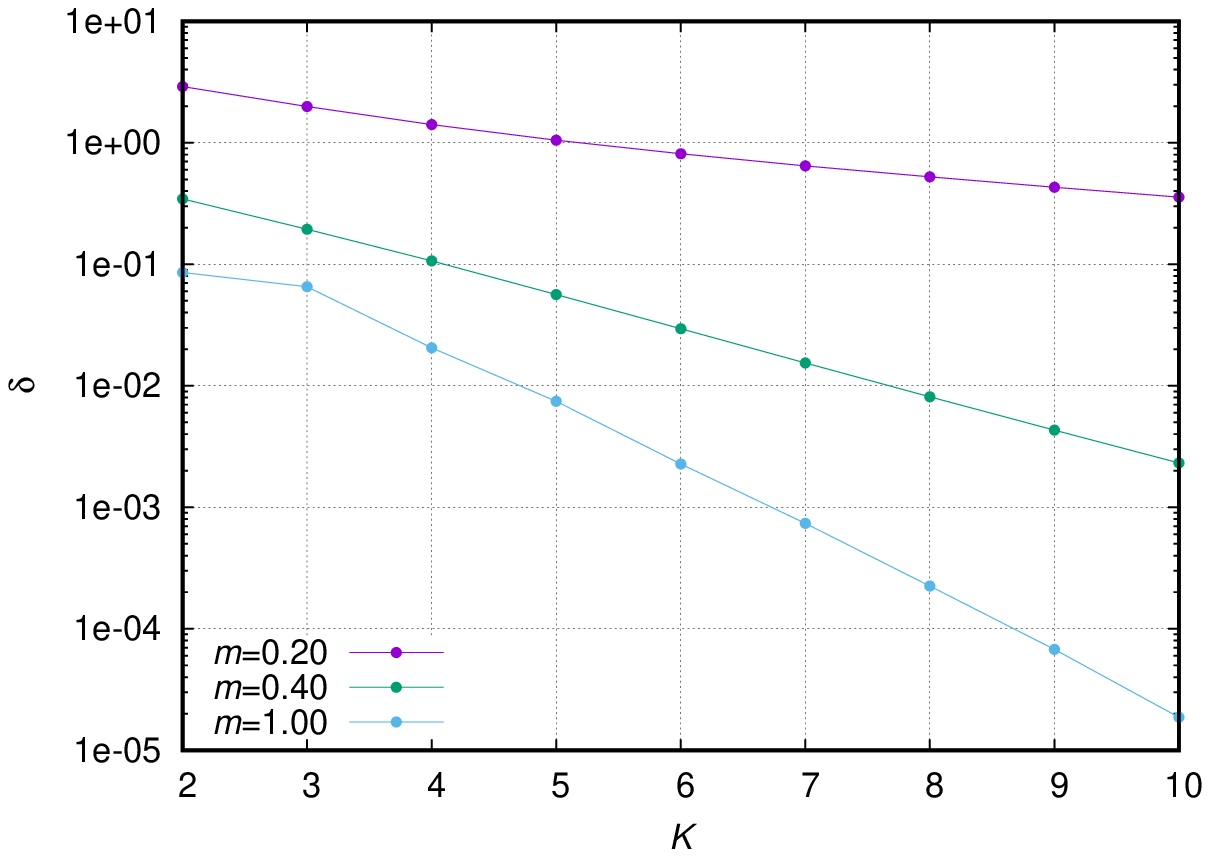}
  \caption{$K$-dependence of the relative errors of $\ln \left(Z_{\rm B}\left(K\right)\right)$ on $V=2 \times 2$ lattice.}
  \label{fig:K-dependence_ZB}
\end{figure}

The $K$-dependence of the relative errors is investigated in figure~\ref{fig:K-dependence_ZB}.
In order to purely see the discretization effect due to finite $K$, we set the maximum bond dimension of the tensor $K^2$
and choose the lattice size $V=2\times2$ that allows us to carry out a full contraction for the computation of the partition function.
Although there are no other systematic errors except for finite $K$, the value of $K$ is practically restricted up to $10$.
Figure~\ref{fig:K-dependence_ZB} shows that the errors decrease by increasing $K$. 
From this we can say that a simple discretization scheme such
as the Gauss–Hermite quadrature well approximates the original integrals if $K$ is sufficiently large,
and that the tensor network representation reproduces the correct values
of the boson partition function.

%
%

\subsection{Witten index of the free $\mathcal{N}=1$ Wess--Zumino model}
\label{sec:Results_freeWZ}

\begin{figure}[htbp]
  \centering
  \includegraphics[width=0.6\hsize]{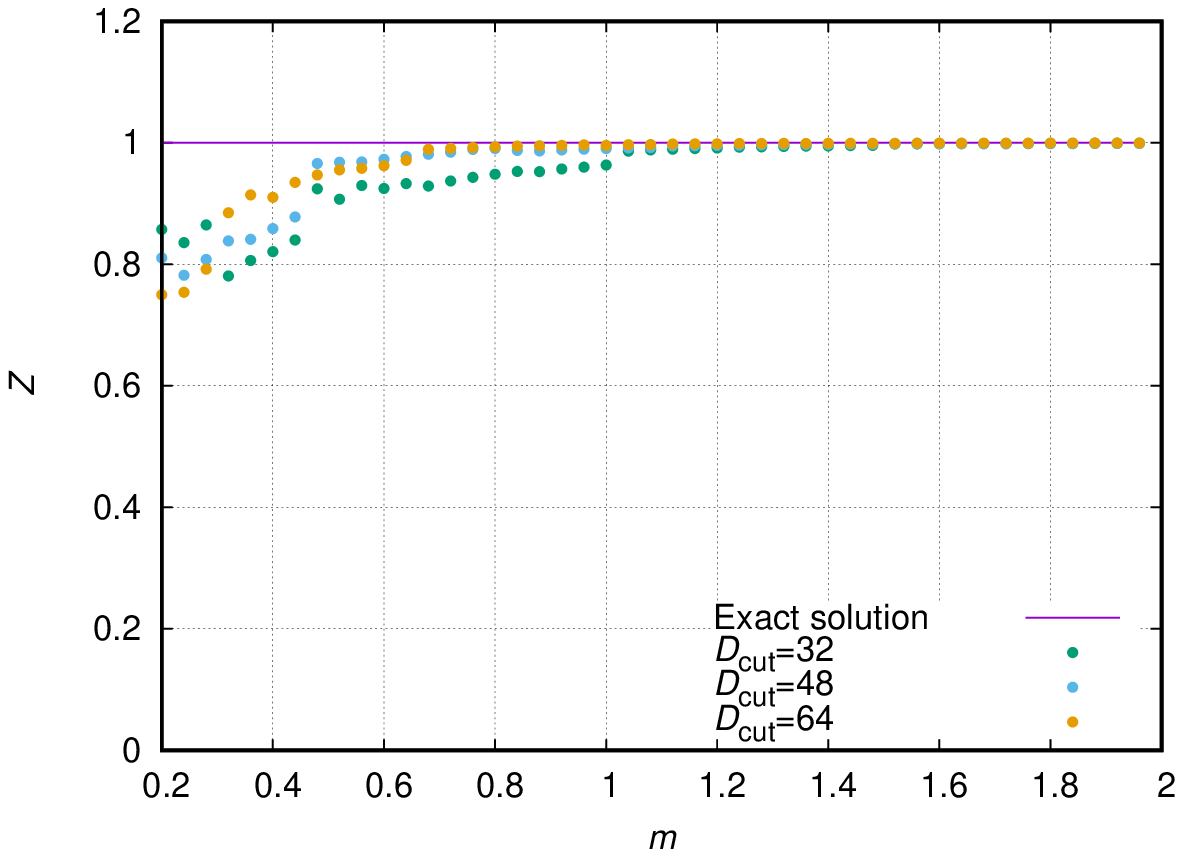}
  \includegraphics[width=0.6\hsize]{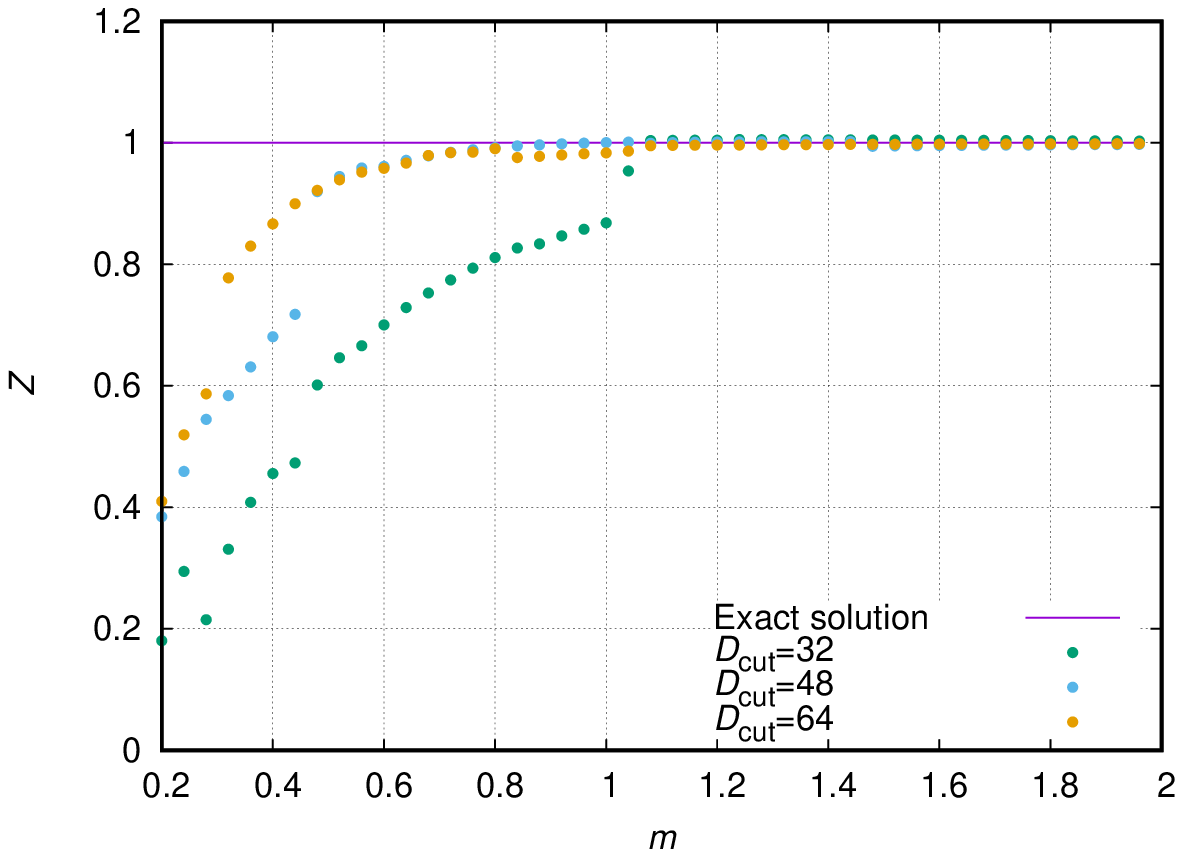}
  \includegraphics[width=0.6\hsize]{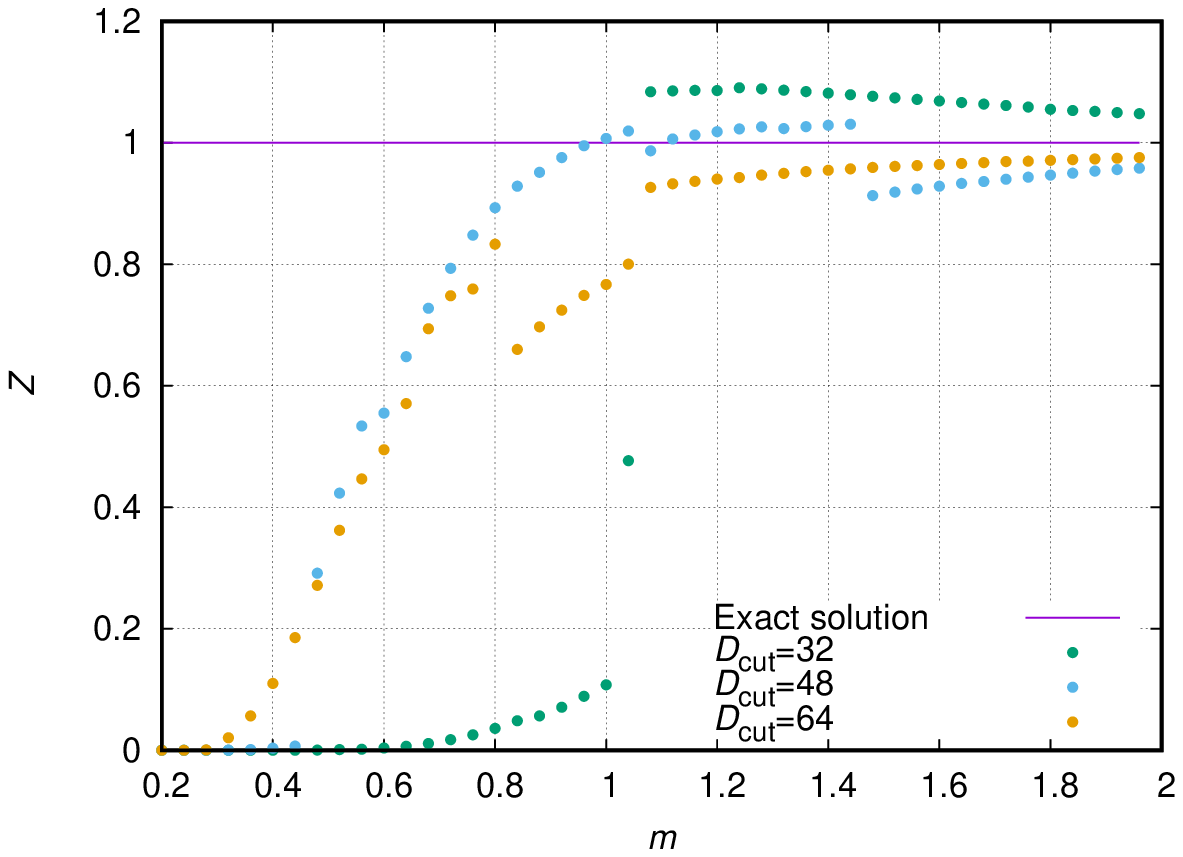}
  \caption{The Witten index of the free Wess--Zumino model against $m$ on $V=2 \times 2$~(top), $8\times 8$~(center), $32 \times 32$~(bottom).
    \label{fig:Witten_index}}
\end{figure}

\begin{figure}[htbp]
  \centering
  \includegraphics[width=0.6\hsize]{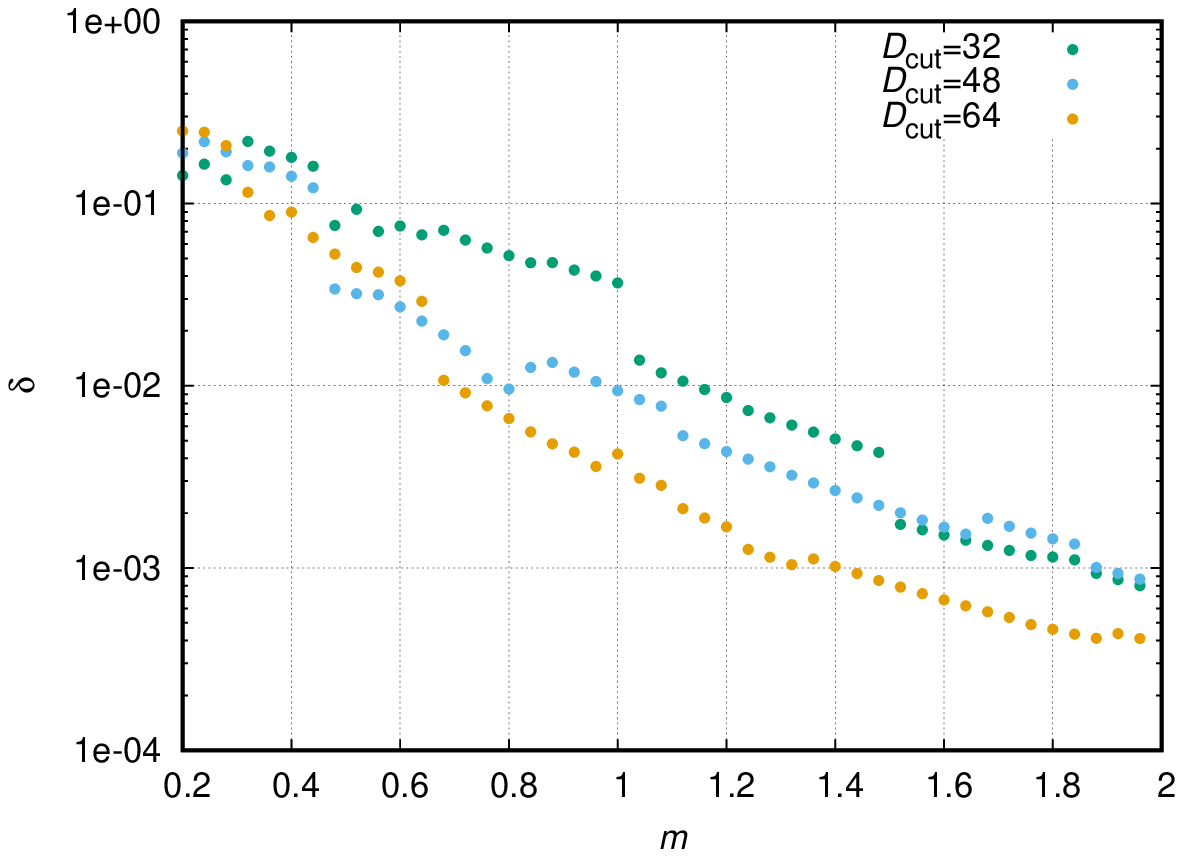}
  \includegraphics[width=0.6\hsize]{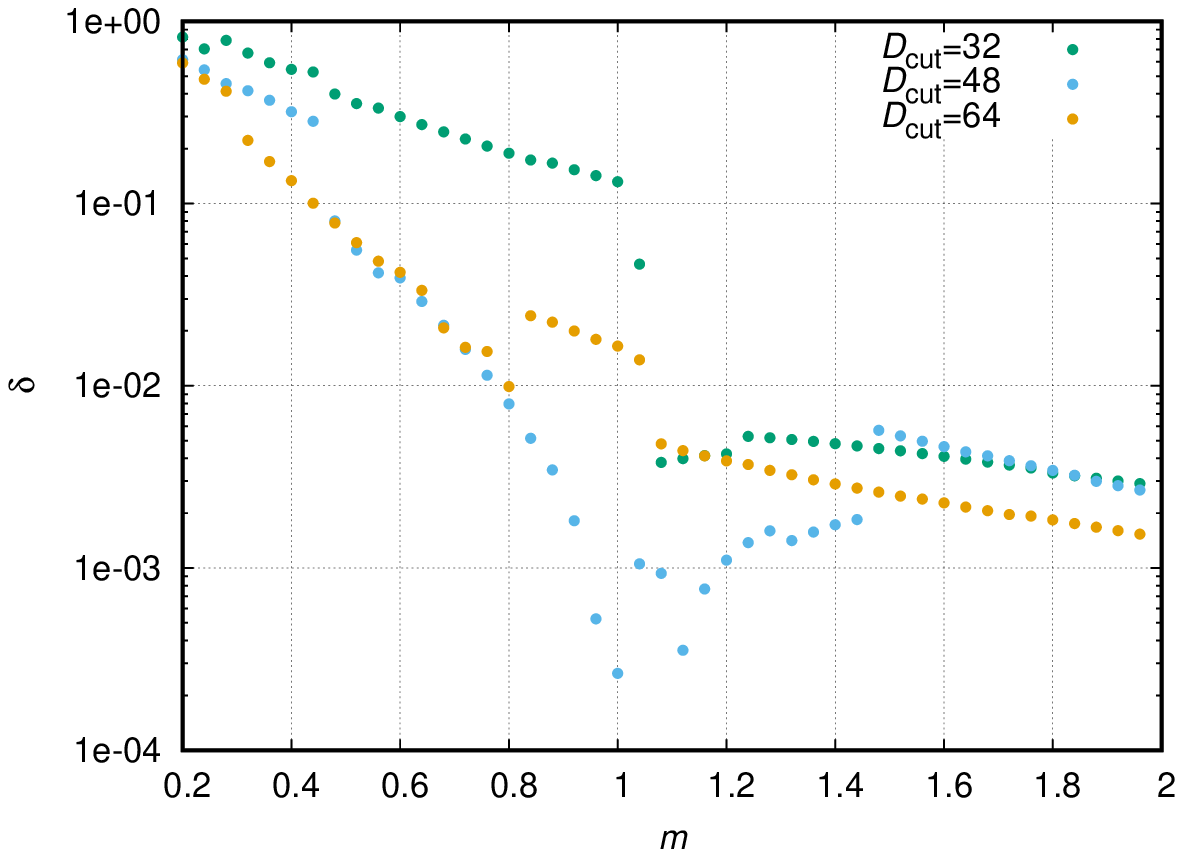}
  \includegraphics[width=0.6\hsize]{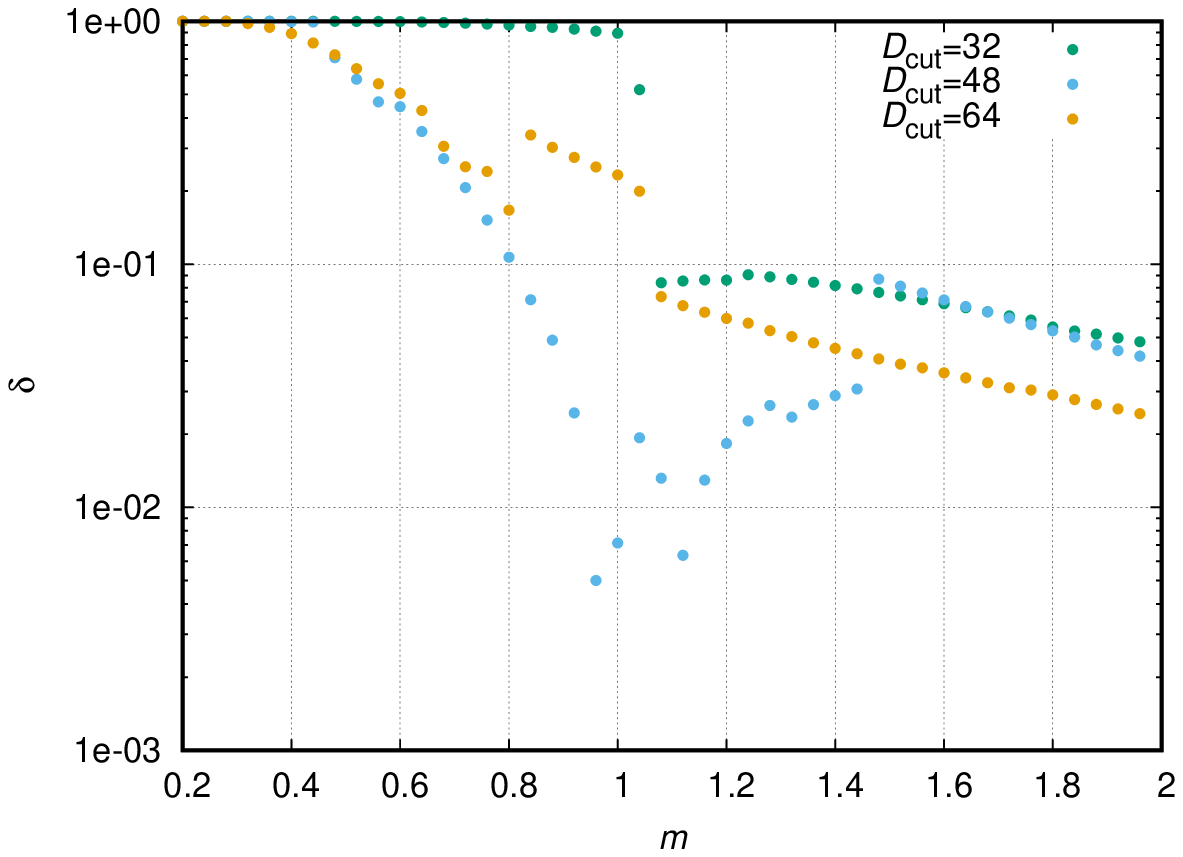}
  \caption{Relative errors of the Witten index as a function of $m$ on $V=2 \times 2$~(top), $8\times 8$~(center), $32 \times 32$ lattices~(bottom).
    \label{fig:Witten_index_error}}
\end{figure}

The Witten index computed by the Grassmann TRG is shown in figure~\ref{fig:Witten_index}.
Figure~\ref{fig:Witten_index_error} shows the relative error of the Witten index.
As discussed in section~\ref{sec:lattice_theory}, the fermion and the boson are decoupled from each other in the free case.
In this section, however, we treat the free Wess--Zumino model as a combined system of fermions and bosons;
thus we perform the Grassmann TRG for a single tensor network.
One can see that the results tend to converge to the exact values by increasing $D_{\rm cut}$.
The obtained indices with $D_{\rm cut}=64$ (yellow symbols) take the values near one
compared with those of $D_{\rm cut}=32$ (green symbols).

Thus we can conclude that eq.~\eqref{eq:19} gives a correct tensor network representation 
of the two-dimensional lattice $\mathcal{N}=1$ Wess--Zumino model.
$Z_{\rm F}$ and $Z_{\rm B}$ become extremely large and extremely small, respectively, for large space-time volume.  
For instance,  $Z_{\rm F}$ are of the order of $\mathcal{O}\left(10^{400}\right)$
at $m=1$ on $V=32 \times 32$ lattice as seen in figure~\ref{fig:ZF}.
Surprisingly, $\mathcal{O}\left(1\right)$ values are obtained as the Witten index
as seen in figure~\ref{fig:Witten_index}.
Namely, the boson effect balancing huge $Z_{\rm F}$ 
is correctly reproduced using the Grassmann TRG for the total tensor.
So we can say that the TRG is a very promising approach 
to study the supersymmetric field theories.

%
%
%
%

\section{Summary and outlook}
\label{sec:Summary}

We have shown that the two-dimensional lattice $\mathcal{N}=1$ Wess--Zumino model is expressed as a tensor network.
The known techniques of making a tensor were refined in the fermion sector
and generalized in the boson sector in the sense that
it is possible to define a tensor for any way of discretizing the integrals for scalar fields.
We have also tested our formulation in the free theory by estimating the Witten index and comparing it with the exact solution. 
The resulting indices reproduce the exact one as $D_{\rm cut}$, the dimension of the truncated tensor indices in the TRG, increases.

Now we are tackling the issue on the supersymmetry breaking by estimating correlation functions from the tensor network. 
Before investigating the physical breaking effects, 
we have to show that the artificial ones by the lattice cut-off disappears in the continuum limit beyond the arguments of the perturbation theory.
We will estimate the expectation value of the action, the supersymmetric Ward--Takahashi identity, and the mass spectra of fermions and bosons to show it. 
We will then see the supersymmetry breaking in the model with the double-well potential by estimating several physical quantities 
and study the phase structure in detail.

Although we have only dealt with the Wilson type discretization of derivatives,
one may use another way such as the domain wall discretization.
In that case, partition functions or Green's functions will be represented as three dimensional tensor networks.
For such higher dimensional tensor networks, the higher order TRG was introduced in Ref.~\cite{2012PhRvB..86d5139X},
and the Grassmann version was also proposed in Ref.~\cite{Sakai:2017jwp}.
In this way one can in principle go this direction; however, the computational cost could be severe.
Therefore further improvements of the algorithm might be needed for the actual computation in higher dimensions.

We emphasize that the methodology of constructing the tensor is given for any superpotential, that is, any interacting case, in this paper.
Since the Wess--Zumino model consists of various building blocks: the scalar field, the Majorana fermion, and their interactions such as the Yukawa- and the $\phi^4$-interactions,
we expect that our method could be very useful in TRG studies of other theories.

\appendix

%
%
%
%

\section{Coarse-graining step in Grassmann TRG}
\label{sec:Coarse-graining_GTRG}

In this appendix, we describe the coarse-graining step in the Grassmann TRG 
for the current boson-fermion system. 
We basically follow ref.~\cite{Takeda:2014vwa}, which deals with a pure fermionic model (the $N_{\mathrm{f}}=1$ Gross--Neveu model)
and show the method in our notation making the difference that comes from the boson part clear.

We begin with the partition function that initially takes the following form:\footnote{
  Although $Z$ and $\mathcal{T}$ depends on $K$ as eqs.~\eqref{eq:19} and~\eqref{eq:total_T}, $K$ is simply abbreviated here.
}
\begin{align}
  \label{initial_form}
  Z= \sum_{\left\{X,T\right\}} &\prod_{n \in \Gamma} \mathcal{T}_{X_{n} T_{n} X_{n-\hat{1}} T_{n-\hat{2}}}
                                 \nonumber \\
                               &\cdot  \int \prod_{n \in \Gamma} \mathrm{d} \Xi_{n}^{uvpq} \cdot
                                 \prod_{n \in \Gamma} 
                                 \left(\bar \xi_{n+\hat 1}\xi_n\right)^{u_n}  \left(\bar \chi_{n+\hat 1}\chi_n\right)^{v_n}  \left(\bar \eta_{n+\hat 2}\eta_n\right)^{p_n}  \left(\bar \zeta_{n+\hat 2}\zeta_n\right)^{q_n},
\end{align}
where the local measure of Grassmann variables is defined as
\begin{align}
  \label{fermion_measure}
  \mathrm{d}\Xi_{n}^{uvpq}=  \mathrm{d}\xi^{u_n}_n \mathrm{d}\chi^{v_n}_n \mathrm{d}\eta^{p_n}_n \mathrm{d}\zeta^{q_n}_n 
  \mathrm{d} \bar \xi^{u_{n-\hat 1}}_n \mathrm{d}\bar \chi^{v_{n-\hat 1}}_n \mathrm{d} \bar \eta^{p_{n-\hat 2}}_{n} \mathrm{d} \bar \zeta^{q_{n-\hat 2}}_{n},
\end{align}
and the tensor elements are not zeros only when
\begin{align}
  \label{eq:12}
  \left(u_{n} + v_{n} + p_{n} + q_{n} + u_{n-\hat{1}} + v_{n-\hat{1}} + p_{n-\hat{2}} + q_{n-\hat{2}}\right) \bmod 2 = 0
\end{align}
holds.
The tensor $\cal T$ is made of the fermionic one 
${T_{\rm F}}\left(\phi_n\right)_{u_n v_n p_n q_n u_{n-\hat 1} v_{n-\hat 1} p_{n-\hat 2} q_{n-\hat 2}}$ in  eq.~\eqref{eq:fermion_tensor}
and the boson one ${T_{\rm B}}\left(K\right)_{w_n s_n w_{n-\hat 1} s_{n-\hat 2}} $ in  eq.~\eqref{eq:boson_tensor}
as in eq.~\eqref{eq:total_T}. The indices $u_n,v_n,p_n,q_n$ take two values $0$ or $1$
while $w_n,s_n$ run from $1$ to $D_{\mathrm{init}}$ as seen in section~\ref{sec:Some_details}.
The total indices $X_n$ and $T_n$ are given by $X_n = \left(u_n,v_n,w_n\right)$ and $T_n = \left(p_n,q_n,s_n\right)$, and they run from 1 to $2 \times 2 \times D_{\mathrm{init}}$.
As mentioned in section~\ref{sec:Some_details},  we set $D_{\mathrm{init}}=D_{\mathrm{cut}}/2$ for the actual computations.

\begin{figure}[htbp]
  \centering
  \includegraphics[width=0.9\hsize]{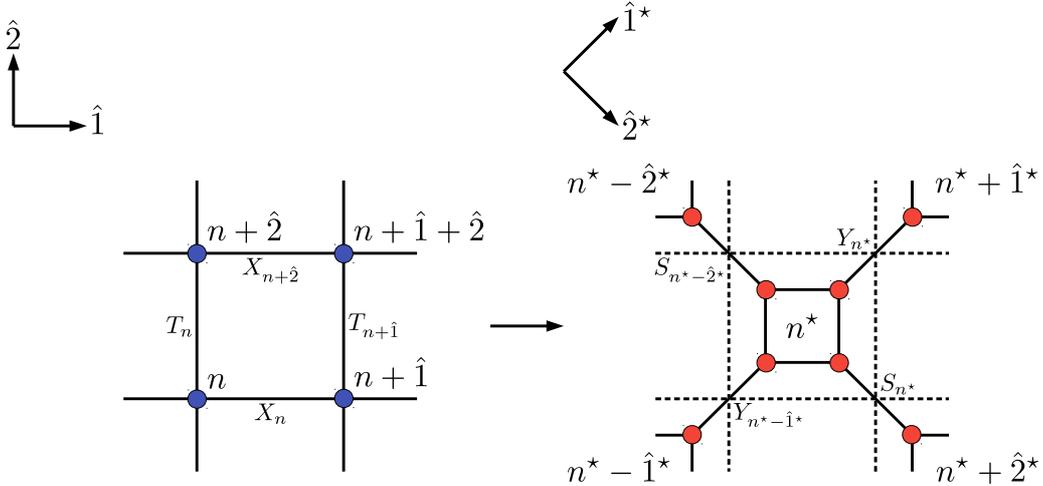}
  \caption{
    Old and new lattice coordinates $n$ and $n^{\star}$.
    The old tensor indices $X, T$ and new ones $S, Y$ are also shown.
    The blue symbols represent the tensors in the RHS of eq.~\eqref{initial_form},
    and the red ones represent the decomposed rank-three tensors appear in the following paragraphs.
  }
  \label{fig:n_and_nstar}
\end{figure}

The coarse-graining of a tensor network mainly consists of three steps:
the SVD of tensors, a decomposition of Grassmann measures, 
and a contraction of the indices and taking the integrals of Grassmann variables defined on $\Gamma$.
The SVD and the decomposition for Grassmann measures are performed in a different manner
for even and odd sites.
We will see that the coarse-grained tensors take the same form as~\eqref{initial_form} with $v=q=0$ and
are defined on the coarse-grained lattice
\begin{align}
  \Gamma^\star= \Set{n+\frac{1}{2}\left(\hat 1 + \hat 2\right)| n=\left(n_1,n_2\right) \in \Gamma, \ \text{where } n_1+n_2 \text{ is an even integer.}}. 
\end{align}
This means that $\Gamma^\star$ is a set of the center of the plaquette
($n,n+\hat1,n+\hat 1+\hat 2,n+\hat 2$) with even sites $n$. The unit vectors of $\Gamma^\star$ 
are $\hat{1}^{\star} = \hat{1} + \hat{2}$ and $\hat{2}^{\star} = \hat{1} - \hat{2}$.
The correspondence between $n$ and $n^{\star}$ is shown in figure~\ref{fig:n_and_nstar}.

First, on even sites $n\in \Gamma$, we just take the truncated SVD of  $\mathcal{T}$ like eq.~\eqref{eq:cg_tensor}:
\begin{align}
  \label{eq:13}
  \mathcal{T}_{X_{n}T_{n}X_{n-\hat{1}}T_{n-\hat{2}}}
  \approx \sum_{w_{n^{\star}-\hat{1}^{\star}} = 1}^{D_{\mathrm{cut}}} U^{1}_{(X_{n}T_{n})w_{n^{\star}-\hat{1}^{\star}}} \sigma^{13}_{w_{n^{\star}-\hat{1}^{\star}}} V^{3\dagger}_{w_{n^{\star}-\hat{1^{\star}}}(X_{n-\hat{1}}T_{n-\hat{2}})},
\end{align}
where
\begin{align}
  \label{eq:8}
  n^\star= n+\frac{1}{2}\left(\hat 1+\hat 2\right) \in \Gamma^\star.
\end{align}
The Grassmann measures are divided into two pieces as
\begin{align}
  \label{eq:17}
  \mathrm{d} \Xi_{n}^{uvpq} = \int
  \left(\Theta^{1}_{n,u_{n}v_{n}p_{n}q_{n}} \mathrm{d}\bar{\xi}_{n^{\star}}^{u_{n^{\star}-\hat{1}^{\star}}}\right)
  \left(\Theta^{3}_{n,u_{n-\hat{1}}v_{n-\hat{1}}p_{n-\hat{2}}q_{n-\hat{2}}} \mathrm{d}\xi_{n^{\star} - \hat{1}^{\star}}^{u_{n^{\star}-\hat{1}^{\star}}}\right)
  \left(\bar{\xi}_{n^{\star}}\xi_{n^{\star}-\hat{1}^{\star}}\right)^{u_{n^{\star}-\hat{1}^{\star}}},
\end{align}
where
\begin{align}
  \label{eq:21}
  &\Theta^{1}_{n,abcd}
    = \mathrm{d}\xi_{n}^{a} \mathrm{d}\chi_{n}^{b} \mathrm{d}\eta_{n}^{c} \mathrm{d}\zeta_{n}^{d},  \\
  &\Theta^{3}_{n,abcd}
    = \mathrm{d}\bar{\xi}_{n}^{a} \mathrm{d}\bar{\chi}_{n}^{b} \mathrm{d}\bar{\eta}_{n}^{c} \mathrm{d}\bar{\zeta}_{n}^{d},
\end{align}
and the new index $u_{n^{\star} - \hat{1}^{\star}}$ is defined as
\begin{align}
  \label{eq:22}
  u_{n^{\star} - \hat{1}^{\star}}
  \equiv \left(u_{n} + v_{n} + p_{n} + q_{n}\right) \bmod 2.
\end{align}
Note that each parenthesized factors on the RHS of eq.~\eqref{eq:17} are Grassmann-even under eqs.~\eqref{eq:12} and~\eqref{eq:22},
and one can freely move them to make a new tensor.
The tensor in eq.~\eqref{eq:13} and the measures in eq.~\eqref{eq:17}  have been decomposed into ($X_{n}T_{n}$)-part 
and ($X_{n-\hat{1}}T_{n-\hat{2}}$)-part,
and they are connected via the new indices ($u_{n^{\star}-\hat{1}^{\star}}, w_{n^{\star}-\hat{1}^{\star}}$).

For odd lattice sites $n+\hat{2}$ next to even sites $n$, we take another decomposition:
\begin{align}
  \label{eq:23}
  \mathcal{T}_{X_{n+\hat{2}}T_{n+\hat{2}}X_{n-\hat{1}+\hat{2}}T_{n}} 
  \approx \sum_{s_{n^{\star}-\hat{2}^{\star}}=1}^{D_{\mathrm{cut}}} U^{2}_{(T_{n}X_{n+\hat{2}})s_{n^{\star}-\hat{2}^{\star}}} \sigma^{24}_{s_{n^{\star}-\hat{2}^{\star}}} V^{4\dagger}_{s_{n^{\star}-\hat{2}^{\star}}(T_{n+\hat{2}}X_{n-\hat{1}+\hat{2}})}.
\end{align}
The Grassmann measure is also decomposed into ($T_{n}X_{n+\hat{2}}$)-part and ($T_{n+\hat{2}}X_{n-\hat{1}+\hat{2}}$)-part as
\begin{align}
  \label{eq:27}
  \mathrm{d} \Xi_{n+\hat 2}^{uvpq}
  = &\int \left(\Theta^{2}_{{n+\hat 2}, p_{n}q_{n}u_{n+\hat{2}}v_{n+\hat{2}}} \mathrm{d}\bar{\eta}_{n^{\star}}^{p_{n^{\star}-\hat{2}^{\star}}}\right)
      \left(\Theta^{4}_{{n+\hat 2}, p_{n+\hat{2}}q_{n+\hat{2}}u_{n-\hat{1}+\hat{2}}v_{n-\hat{1}+\hat{2}}} \mathrm{d}\eta_{n^{\star} - \hat{2}^{\star}}^{p_{n^{\star}-\hat{2}^{\star}}}\right)
      \nonumber \\
    & \cdot \left(\bar{\eta}_{n^{\star}}\eta_{n^{\star}-\hat{2}^{\star}}\right)^{p_{n^{\star}-\hat{2}^{\star}}},
\end{align}
where
\begin{align}
  \label{eq:28}
  &\Theta^{2}_{n,abcd}
    = \left(-1\right)^{a + b}  \mathrm{d}\bar{\eta}_{n}^{a} \mathrm{d}\bar{\zeta}_{n}^{b} \mathrm{d}\xi_{n}^{c}
    \mathrm{d}\chi_{n}^{d}, \\
  &\Theta^{4}_{n, abcd}
    = \mathrm{d}\eta_{n}^{a} \mathrm{d}\zeta_{n}^{b} \mathrm{d}\bar{\xi}_{n}^{c} \mathrm{d}\bar{\chi}_{n}^{d},
\end{align}
and $p_{n^{\star}-\hat{2}^{\star}}$ is defined by
\begin{align}
  \label{eq:29}
  &p_{n^{\star} - \hat{2}^{\star}}
    \equiv \left(p_{n} + q_{n} + u_{n+\hat{2}} + v_{n+\hat{2}}\right) \bmod 2.
\end{align}
Note that the extra sign in eq.~\eqref{eq:28} arise from the rearrangement of the Grassmann measures.

We thus find that the partition function can be expressed in terms of coarse-grained tensor:
\begin{align}
  \label{eq:32}
  Z  \approx \sum_{\left\{Y, S\right\}} &\prod_{n^{\star} \in \Gamma^{\star}} 
                                          \mathcal{T}^{\rm new}_{Y_{n^{\star}} S_{n^{\star}} Y_{n^{\star}-\hat{1^{\star}}} S_{n^{\star}-\hat{2^{\star}}}} \nonumber\\
                                        & \cdot \int \prod_{n^{\star} \in \Gamma^{\star}} \mathrm{d} \Xi^{up}_{n^\star}
                                          \prod_{n^{\star} \in \Gamma^{\star}} \left(\bar{\xi}_{n^{\star}+\hat{1}^{\star}}\xi_{n^{\star}}\right)^{u_{n^{\star}}} \left(\bar{\eta}_{n^{\star}+\hat{2}^{\star}}\eta_{n^{\star}}\right)^{p_{n^{\star}}}
\end{align}
with
\begin{align}
  \label{eq:33}
  \mathrm{d} \Xi^{up}_{n^\star}
  = \mathrm{d}\xi_{n^{\star}}^{u_{n^{\star}}} \mathrm{d}\eta_{n^{\star}}^{p_{n^{\star}}} \mathrm{d}\bar{\xi}_{n^{\star}}^{u_{n^{\star}-\hat{1}^{\star}}} \mathrm{d}\bar{\eta}_{n^{\star}}^{p_{n^{\star}-\hat{2}^{\star}}},
\end{align}
where the coarse-grained tensor with new indices $Y_{n^{\star}} = \left(u_{n^{\star}}, w_{n^{\star}}\right)$ and $S_{n^{\star}} = \left(p_{n^{\star}}, s_{n^{\star}}\right)$ 
is defined  by
\begin{align}
  \label{eq:30}
  &\hspace{-0.8em}\mathcal{T}^{\rm new}_{Y_{n^{\star}} S_{n^{\star}} Y_{n^{\star}-\hat{1^{\star}}} S_{n^{\star}-\hat{2^{\star}}}} \nonumber\\
  = &\sqrt{\sigma^{13}_{w_{n^{\star}}} \sigma^{24}_{s_{n^{\star}}} \sigma^{13}_{w_{n^{\star}-\hat{1}^{\star}}} \sigma^{24}_{s_{n^{\star}-\hat{2}^{\star}}}}
      \nonumber \\
  & \cdot \sum_{X_n} 
    \sum_{T_n} 
    \sum_{X_{n+\hat 2}} 
    \sum_{T_{n+\hat 1}} 
    U^{1}_{(X_{n}T_{n})w_{n^{\star}-\hat{1}^{\star}}}
    U^{2}_{(T_{n}X_{n+\hat{2}})s_{n^{\star}-\hat{2}^{\star}}}
    V^{3\dagger}_{w_{n^{\star}}(X_{n+\hat{2}}T_{n+\hat{1}})}
    V^{4\dagger}_{s_{n^{\star}}(T_{n+\hat{1}}X_{n})} 
    \nonumber \\
  & \cdot \begin{aligned}[t]
    \int 
    &\Theta^{2}_{n+\hat 2, p_{n}q_{n}u_{n+\hat{2}}v_{n+\hat{2}}}
    \Theta^{1}_{n,u_{n}v_{n}p_{n}q_{n}} 
    \Theta^{4}_{n+\hat 1, p_{n+\hat{1}}q_{n+\hat{1}}u_{n}v_{n}}
    \Theta^{3}_{n+\hat 1+\hat 2, u_{n+\hat{2}}v_{n+\hat{2}}p_{n+\hat{1}}q_{n+\hat{1}}} 
    \nonumber \\
    & \cdot 
    \left(\bar{\xi}_{n+\hat{1}}\xi_{n}\right)^{u_{n}} 
    \left(\bar{\chi}_{n+\hat{1}}\chi_{n}\right)^{v_{n}} 
    \left(\bar{\eta}_{n+\hat{2}}\eta_{n}\right)^{p_{n}} 
    \left(\bar{\zeta}_{n+\hat{2}}\zeta_{n}\right)^{q_{n}} 
    \nonumber \\
    & \cdot 
    \left(\bar{\xi}_{n+\hat{1}+\hat{2}}\xi_{n+\hat{2}}\right)^{u_{n+\hat{2}}} 
    \left(\bar{\chi}_{n+\hat{1}+\hat{2}}\chi_{n+\hat{2}}\right)^{v_{n+\hat{2}}} 
    \left(\bar{\eta}_{n+\hat{1}+\hat{2}}\eta_{n+\hat{1}}\right)^{p_{n+\hat{1}}}       
    \left(\bar{\zeta}_{n+\hat{1}+\hat{2}}\zeta_{n+\hat{1}}\right)^{q_{n+\hat{1}}}
  \end{aligned}
      \nonumber \\
  & \cdot \delta_{(u_{n+\hat{2}}+v_{n+\hat{2}}+p_{n+\hat{1}}+q_{n+\hat{1}})\bmod 2, u_{n^{\star}}} 
    \delta_{(p_{n+\hat{1}}+q_{n+\hat{1}}+u_{n}+v_{n})\bmod 2, p_{n^{\star}}} 
    \nonumber \\
  & \cdot \delta_{(u_{n}+v_{n}+p_{n}+q_{n}) \bmod 2, u_{n^{\star}-\hat{1}^{\star}}} 
    \delta_{(p_{n}+q_{n}+u_{n+\hat{2}}+v_{n+\hat{2}}) \bmod 2, p_{n^{\star}-\hat{2}^{\star}}}.
\end{align}
The constraints described in eqs.~\eqref{eq:22} and~\eqref{eq:29} with eq.~\eqref{eq:12} are explicitly imposed as Kronecker deltas.

Owing to the similarity of the initial tensor and the resulting one, the procedure described in this appendix can be simply iterated by setting eq.~\eqref{eq:30} as an initial tensor for the next coarse-graining step.\footnote{
  Strictly speaking, one has to regard that the new coordinates on $\Gamma^{*}$ is expressed by integers as was the case with old ones on $\Gamma$.
}
An important change is the absence of $\chi$ and $\zeta$, so one has to also set $v_{n}$ and $q_{n}$ to 0 for the following steps.
Equation~\eqref{eq:30} and the initial tensor have the different contents of indices,
e.g. $X_{n} = \left(u_{n}, v_{n}, w_{n}\right)$ reduces to $Y_{n^{\star}} = \left(u_{n^{\star}}, w_{n^{\star}}\right)$ after the coarse-graining step.
This means that the dimension of the tensor indices changes from $2\times 2 \times D_{\mathrm{init}}$ to $2 \times D_{\mathrm{cut}}$.
We take $D_{\mathrm{init}} = D_{\mathrm{cut}}/2$ in section~\ref{sec:Results_freeWZ} to retain the size of tensors for the sake of simplicity.

Note also that the definition of the unit vectors turns out to be proportional to original ones after the next coarse-graining step, i.e. $\hat{1}^{\star\star} = \hat{1}^{\star} + \hat{2}^{\star} = 2\cdot \hat{1}$ and $\hat{2}^{\star\star} = \hat{1}^{\star} - \hat{2}^{\star} = 2\cdot \hat{2}$.
Although the coarse-grained lattice $\Gamma^{\star}$ is not isotropic and the boundary conditions are not the same as original ones, 
this strange situation will recover after the next coarse-graining step (see ref.~\cite{Takeda:2014vwa}).

\acknowledgments

We thank Dr.~Yuya Shimizu for his many helpful comments.
This work is supported in part by
JSPS KAKENHI Grant Numbers JP16K05328, JP17K05411,
Grants-in-Aid for Scientific Research from the Ministry of Education, Culture, Sports, Science and Technology (MEXT) (No. 15H03651),
MEXT as ``Exploratory Challenge on Post-K computer (Frontiers of Basic Science: Challenging the Limits)'',
and the MEXT-Supported Program for the Strategic Research Foundation at Private Universities Topological Science (Grant No. S1511006).

\providecommand{\href}[2]{#2}\begingroup\raggedright\endgroup

\end{document}